\documentclass[aps,prx,twocolumn,showkeys,floatfix,nobalancelastpage]{revtex4-2}
\usepackage{comment}
\usepackage{braket}
\usepackage{graphicx}
\usepackage{dcolumn}
\usepackage{bm}
\usepackage{siunitx}
\usepackage{xcolor} 
\usepackage{tabularx}
\usepackage{dsfont}	
\usepackage{mathrsfs} 
\usepackage{float}
\usepackage{ragged2e}
\usepackage{amsmath}
\usepackage{amssymb}
\usepackage{physics}
\usepackage{minitoc}
\usepackage{soul}
\usepackage{csquotes}
\usepackage{tikz}
\usetikzlibrary{shapes.geometric, positioning, decorations.text}
\usepackage{hyperref}

\usepackage{lineno}
\usepackage{graphicx}
\usepackage{tikz}
\usetikzlibrary{positioning, arrows.meta}
\usepackage{xcolor}
\usepackage{soul}
\newcommand{\exchangeterm}{\mathcal{M}
_{n,m}} 

\definecolor{hotpink}{RGB}{170,40,120}
\definecolor{darkred}{HTML}{A11D1D}
\definecolor{darkorange}{HTML}{C45A00}

\newcommand{\Novelty}[1]{{\color{darkorange}#1}}
\newcommand{\Contribution}[1]{{\color{hotpink}#1}}
\newcommand{\ReviewerComment}[2]{{\color{darkred}
~#2}}

\renewcommand{\Novelty}[1]{#1}
\renewcommand{\Contribution}[1]{#1}
\renewcommand{\ReviewerComment}[2]{#2}

\begin{document}
\raggedbottom

\title{
AN EXTENSIVE THEORY OF NONLINEARLY INTERCOUPLED \\ PSEUDOMODES FOR NOISE MODEL REDUCTION IN CIRCUIT QED
}
\author{{M. Gabriela Boada G.}\textsuperscript{1}}
    \email[Correspondence email address:]{maria.boada@utsa.edu}
\author{{Nicolas Dirnegger}\textsuperscript{2}}
\author{{Andrea Delgado}\textsuperscript{4}}
\author{Prineha Narang\textsuperscript{2,3}}
    \affiliation{\textsuperscript{1}University of Texas, San Antonio, Dept. of Physics \& Astronomy, San Antonio, TX 78249, USA}
    \affiliation{\textsuperscript{2}Department of Electrical and Computer Engineering, UCLA, Los Angeles, CA 90095, USA}
    \affiliation{\textsuperscript{3}College of Letters and Science, University of California, Los Angeles, CA 90095, USA}
    \affiliation{\textsuperscript{4} Qblox B.V., Delft, 2628 CJ, The Netherlands.}
\date{\today}

\newcommand{\Redbold}[1]{{\color{black}#1}}
\newcommand{\Clarity}[1]{{\color{black}#1}}
\newcommand{\Connection}[1]{{\color{black}#1}}

\begin{abstract}
Superconducting circuit quantum electrodynamics (cQED) platforms present a persistent modeling challenge: the intrinsic nonlinearity of the Josephson potential couples to a dissipative electromagnetic environment in ways that resist both perturbative treatment and Markovian reduction.~Standard approaches either scale poorly with system size or rely on approximations about the noise structure whose validity is regime-dependent and rarely verified at the level of the modeled hardware.~In this work, we extend Garraway's generalized pseudomode \cite{Garraway1997} construction to accommodate nonlinearly intercoupled principal subsystem ensembles of the form of the Kerr-cat Hamiltonian, providing a nonperturbative framework for circuit QED noise model reduction well-suited for isolating the dynamics of the mode of interest.~We develop the general theory in the Heisenberg picture, then demonstrate an open-form elimination for two-, three-, and four-mode Kerr-coupled systems with bilinear exchange and three-wave mixing interactions, establishing the precise equivalence between a displaced four-mode parent and the effective three-wave-mixing spectrum observed under a stiff, coherent pump.~The construction is intended for structured, strongly driven regimes that resist the standard Markovian treatment.
\end{abstract}

\maketitle
\section{Introduction}\label{sec:introduction}
Superconducting cQED has emerged over the past two decades as one of the leading platforms for quantum information processing, metrology, and many-body physics ~\cite{Wallraff2004,Blais2021}.~The Josephson junction enables the complete toolbox of parametric processes that make modern hardware-efficient encodings possible, from Kerr-cat and dissipative cat qubits~\cite{Frattini2021,Mirrahimi2014,Leghtas2015} to concatenated bosonic codes~\cite{Putterman2025, Mirrahimi2014}.~This same feature is, however, the source of its most persistent modeling challenges~\cite{Willsch2024,Cohen2023,Shillito2022,Venkatraman2022}.~Because the Josephson potential is nonpolynomial in the circuit flux, any honest quantum description of a circuit QED processor must retain nonlinearity at the Hamiltonian level, even before coupling to the environment is considered~\cite{Nigg2012,Minev2021,Malekakhlagh2020}.~Accurately modeling how such nonlinear subsystems interact with their dissipative surroundings is nontrivial.~In practice, the response is to bolt on successive corrections each derived perturbatively and each valid only in a specific regime \cite{Nigg2012, Wallraff2004}.~The conventional workflow produces tractable master equations at the price of two structural assumptions: that the environment is memoryless on all timescales relevant to the dynamics, and that the system-environment coupling is weak enough for Born-Markov to apply~\cite{Breuer2002,GardinerZoller2004,Lindblad1976,Blais2021}.~The resulting models are accurate where they were calibrated and increasingly unreliable everywhere else \cite{Leghtas2015, Carde2025}. 

The open quantum systems community has, in part, developed a mature alternative.~Garraway's pseudomode construction~\cite{Garraway1997,Dalton2001, Pleasance2020} replaces the continuum reservoir with a finite set of discrete (auxiliary) damping modes whose reduced dynamics reproduce the exact memory kernel of the original environment.~Later works generalized the construction to arbitrary rational spectral densities with multi-excitation dynamics,~and non-Hermitian auxiliary modes~\cite{Tamascelli2018,Pleasance2020,Lambert2019,Menczel2024,Mazzola2009}.~The upshot is a nonperturbative, numerically exact reduction:~When the environmental response function is rational in the frequency domain, the environment admits an exact representation by a finite set of damped auxiliary modes whose reduced dynamics reproduce the original memory kernel.~More generally, for non-rational response functions, the same construction can be approximated by fitting either the spectral density or the environmental auto-correlation function 
with a finite rational polar decomposition.

Conversely, the literature in cQED has largely treated environmental coupling through the Markovian assumption described above, precisely the regime in which pseudomodal reductions offer the clearest advantage.~In this work, we show that the reduction of the Kerr-cat Hamiltonian \cite{Nigg2012} does survive, and we develop the framework for obtaining closure explicitly.~The key observation here is that pseudomodal eliminations are not implicitly tied to linearity but to representability:~any eliminated subsystem can be replaced by a finite set of auxiliary modes whenever its influence on the retained subsystem is captured by a well-defined response function.

\Novelty{Although entirely permissible within Pleasance's~\cite{Pleasance2020} general construction, there exist no concrete examples in the literature in which the principal subsystem ensemble is an interacting nonlinear oscillator.~Previous considerations of ensembles have primarily been focused on harmonic modes~\cite{Menczel2024,Liang2024}, whereas nonlinear retained systems have largely been limited to finite-dimensional auxiliary spaces~\cite{Dalton2001,Pleasance2022,Zhang2025}.~To our knowledge, explicit constructions for ensembles of self-interacting Kerr-type oscillators have yet to be considered, leaving the resulting occupation-dependent dynamics of the retained subsystem unexplored.~Extending this method of noise model reduction to the class of Hamiltonians used to describe superconducting circuits is of particular value- the device Hamiltonians generic to circuit QED hardwares~\cite{Nigg2012} require the existence of and nonlinear coupling between multiple modes, whereas only the open-form dynamics of the single information-carrying mode are of concern.~In this work, we do so, and we consider the repercussions for the dynamics of the retained subsystem.~We show how the projected response function of such ensembles admits an occupation-conditioned Green's function description \cite{Guo2020, Du2021FrequencyConversion} whose associated self-energy can be expressed as a hierarchy of rational pole expansions.}

Section~\textbf{~\ref{sec:background}.} overviews Garraway's~\cite{Garraway1997} pseudomode construction and establishes its connection to the retarded Green's-functions, introducing the Dyson-type equation that motivates the explicit construction for Kerr- type nonlinearities introduced in Section~\textbf{\ref{sec:general_theory}}.~Section~\textbf{\ref{sec:mode-elimination}.} develops the two-mode prototype, establishing the universal reduction template in which the subsequent three-mode and four-mode constructions follow by a successive application of the same elimination procedure.~Last, Section\textbf{~\ref{sec:conclusion}.} combines all of the results from the reductions obtained in Section \textbf{\ref{sec:coherentdisplacement}} to demonstrate how coherent displacements~\cite{Wallraff2004, Blais2021} of an eliminated mode can still activate a higher-order mixing channel in the pseudomodal reduction, ensuring the direct applicability of our work towards the optimization of parametric processes in circuit QED, such as frequency conversion, parametric mixing, and driven-dissipative protocols commonly encountered in manufactured dissipation~\cite{Leghtas2015,Murch2012,Shankar2013}.

\section{Background}\label{sec:background}

Garraway's pseudomodes construction~\cite{Garraway1997, Dalton2001, Pleasance2020} replaces an enlarged subsystem-environment problem by a finite set of (auxiliary) degrees of freedom whose reduced dynamics reproduce the memory kernel seen by the retained subsystem at the continuum limit.~{We begin by introducing the two-time reservoir autocorrelations, which completely characterize the reservoir influence,
\begin{equation}
    \langle B_\alpha^{\dagger}(t) B_\beta(s) \rangle_E \equiv
\int_0^\infty d\omega\,
J_{\alpha\beta}^{(\ell)}(\omega)e^{-i\omega(t-s)} .
\end{equation}
where expectation values over the bath oscillator degrees of freedom are denoted with
$
\langle \cdot \rangle_E
\equiv
\mathrm{Tr}_E[\rho_E(\cdot)],
$
where $\rho_E$ is the density operator of the eliminated subsystem and $\mathrm{Tr}_E$ denotes the partial trace over its Hilbert space while $\alpha$ and $\beta$ label independent coupling channels between the retained and eliminated subsystems, with $\{B_\alpha,\,B_\beta\}$ denoting the corresponding environmental operators. 

The quantity $J_{\alpha\beta}^{(\ell)}(\omega)$ is the (cross-)spectral density, describing the frequency-dependent strength of coupling $\ell$ for each of the channels between $\{\alpha\}$ and $\{\beta\}$.~Physically, it specifies how the environmental mode couplings are distributed in frequency and therefore completely determines temporal correlations of the collective system. Its poles, denoted %
$\omega_\ell \leftarrow\!\!\rightarrow \omega_{\alpha \beta} 
$
, are associated with the characteristic frequencies and decay rates of the reservoir oscillators in the Caldeira-Leggett environmental model \cite{Caldeira1983} for Ohmic dissipation we assume throughout.    

At zero temperature, the environment is assumed to occupy its vacuum state, implying that all normally ordered correlations vanish \cite{Gardiner}. In its standard form, Garraway's method rests on the observation that the only vanishing two-time correlation function on $(t \geq s \geq 0)$, for $t'=t-s$ is 
\begin{align}
\label{eq:AoC1}
    \langle B_\alpha^{\dagger}(t) B_\beta(s) \rangle_E &= 0,\quad \forall\, \alpha,\, \beta,\\ \label{eq:AoC2}
    \langle B_\alpha(t) B_\beta(s) \rangle_E &= 0 = \langle B_\alpha^{\dagger}(t) B_\beta^{\dagger}(s) \rangle_E.
\end{align} 

The reservoir autocorrelation functions in Eq.~\eqref{eq:AoC1}-\eqref{eq:AoC2}} may be \textit{replaced} by a finite sum of damped modes, so-called pseudomodes for their appearance in the auxiliary model as the finite and discrete set of $\ell$-many (denoted $N_\ell$) damped modes, indexed as 
\begin{equation}
\label{eq: polecontinuation}
\langle B_\alpha(t) B_\beta^{\dagger}(s) \rangle_E
= \sum_{\alpha,\, \beta}^{N_\ell}\,
g_{\alpha}^{(\ell)} g_{\beta}^{(\ell)} e^{-i\,\omega_{\ell}(t-s)},
\end{equation}
which later serve as the coupling rates for the equivalent \textit{reduced} non-Markovian model.~In the expression above, $g_{\alpha}$ and $g_{\beta}$ denote the coupling amplitudes between the system and bath mode $\ell$ associated with the transitions into $\{\alpha\}$ and $\{\beta\}$, respectively.~For compactness, we define the shorthand %
$
    g_{\alpha\beta}
    \equiv
    g_{\alpha} g_{\beta},
$ %
throughout.

The self-/cross-couplings of the enlarged system are encoded in the Green's function solutions of the Heisenberg Equations of Motion~(EoM) and can be grouped into the reservoir autocorrelations Eq.~(\ref{eq:AoC1})-(\ref{eq:AoC2}),
\begin{equation}
\label{eq: greens}
G_{\alpha \beta}^{(\ell)}(t-s) \equiv \langle B_\alpha(t) B_\beta^{\dagger}(s) \rangle_E.
\end{equation}
For each coupling channel in $\ell$ the Heisenberg EoM define the propagator for the bath oscillators,%
\begin{equation}
B_\alpha(t)
=
\sum_\beta
G_{\alpha\beta}^{(\ell)}(t-s)B_\beta(s),
\quad t\ge s ,
\end{equation}%
where $G_{\alpha\beta}(t-s)$ is the amplitude for an excitation initially in channel $\beta$ at time $s$ to appear in channel $\alpha$ at time $t$. Passing to the frequency domain, 
\begin{align}
\label{eq: greens_polar}
G_{\alpha\beta}^{(\ell)}(t-s)
=
\frac{1}{2\pi i}
\oint_C
\tilde{G}_{\alpha\beta}^{(\ell)}(\omega)
e^{-i\omega(t-s)}
\,d\omega,
\end{align}
the corresponding time-domain Green's function is recovered through the inverse Laplace transform wherein the contour $C$ encloses the poles of the analytically continued response function.~For each retained subsystem, indexed by $\mu$, we define 
\begin{equation}
G_{\alpha\beta}^{(\mu)}(t-s)
=
\mathrm{Tr}_{S\neq \mu}\{
\langle
B_\alpha^{(\mu)}(t)
B_\beta^{(\mu)\dagger}(s)
\rangle_E\},
\end{equation}
where the superscript $(\mu)$ indicates that the environmental operators and Green's function are associated with the environment obtained after eliminating all subsystems except the retained subsystem ($\mu$). Then, we write the channel-resolved Green's function as 
\begin{equation}\tilde G_{\alpha\beta}^{(\mu)}(\omega)
=
-g_{\alpha}^{(\mu)}g_{\beta}^{(\mu)*} J_{\alpha\beta}^{(\mu)}(\omega)
\end{equation}
where $\{g_{\alpha}^{(\mu)},\,g_{\beta}^{(\mu)}\}$ are the new coupling amplitudes connecting the retained subsystem, denoted with the index $\mu$, to the eliminated subsystem response through $\alpha$ and $\beta$.~Then, the prefactors are the pseudomode-resolved couplings obtained by factorizing the residue of the $\ell$th pole,
\begin{equation}
\operatorname*{Res}_{z=z_\ell}
\tilde{G}^{(\mu)}_{\alpha\beta}(z)
=
-g^{(\mu)}_\alpha g^{(\mu)*}_\beta
\lim_{z\to z_\ell}
(z-z_\ell)J^{(\mu)}_{\alpha\beta}(z).
\end{equation}
s.t. $\operatorname{Res}[
\tilde{G}_{\alpha\beta}^{(\mu)}(z)/g_{\alpha}^{(\mu)}g_{\beta}^{(\mu)*}
]\longrightarrow g_{\alpha}^{(\ell)}g_{\beta}^{(\ell)*}$.~When $\tilde{G}_{\alpha\beta}^{(\mu)}(\omega)$ is meromorphic, the uniqueness of the Green's functions guarantees the response function of the (reduced) non-Markovian model uniquely coincides with that of the enlarged Markovian model. By Cauchy's residue theorem,
\begin{align}
\label{eq: residuetheorem}
\frac{1}{2\pi i}\oint_C J_{\alpha\beta}^{(\mu)}(z)\,dz
&=
\sum_{z_\ell\in C}
\operatorname*{Res}_{z=z_\ell}
J_{\alpha\beta}^{(\mu)}(z)
\\&=
\sum_{z_\ell\in C}
\lim_{z\to z_\ell}
(z-z_\ell)\,
J_{\alpha\beta}^{(\mu)}(z).
\end{align}
If $J_{\alpha\beta}\,\!^{(\mu)}
$ is meromorphic~\cite{Economou2006}, its poles can be enclosed in the upper half of the frequency domain, ($t \geq 0$), and can be written as coupling rates in the
same manner as was done for the frequency-dependent reservoir.~The enlarged Markovian system is, therefore,
embedded into the non-Markovian integro-differential equation involving a
finite set of discrete damped modes,
\begin{equation}
    \sum_{\ell}
    g_{\alpha\beta}^{(\ell)} e^{-i\,\omega_{\ell}(t-s)}
    \approxeq \oint_C d\omega\,\tilde{G}_{\alpha\beta}^{(\mu)}(\omega)e^{-i\omega(t-s)}.
\end{equation}

\ReviewerComment{A1}{The principal subsystem may then be eliminated whenever its influence over the retained subsystem can be captured by some well-defined response function, and the resulting polar decomposition of the reduced model closes to the desired level of description.}~This provides a nonperturbative framework for noise model reduction that is exact when the eliminated response is rational and a controlled rational-fit approximation otherwise.~Green's functions then serve as the natural mathematical framework for constructing pseudomodal reductions.~Rather than relying on linearity of the retained subsystem, this approach exploits the analytic structure of the environmental response function, allowing for the memory kernel to be characterized entirely through the poles of the corresponding Green's functions.

\ReviewerComment{A1}{While Pleasance's~\cite{Pleasance2020} generalization of Garraway's pseudomodes~\cite{Garraway1997, GarrawayDalton2006} presumes the simplistic (bi-)linear subsystem-environment coupling in the enlarged Markovian model, preserving the linearity of the finite polar decomposition as we'll do throughout the text, we'll depart from the generalized treatment to consider a specific class of principal subsystem ensembles where the coupling between them is explicitly not linear. Then, elimination of a principal subsystem - even under bilinear coupling - is not inconsequential in the higher-order channels.}

To reflect this kind of exchange in the reduced model, one additional degree of novelty in our treatment 
is that coherent drives are incorporated through an operator-valued field bias, $E(t)$.~With this bath specification, the total bath force, $F(t)$, induces drive-dependent correlations, and the fluctuation-dissipation (FDT) relations encoding the bath memory~\cite{PelargonioZaccone2023, CuiZaccone2018} acquire an explicit dependence~\cite{GambaCuiZaccone2025} set by the two-time autocorrelation of the applied field, $E(t)E(s)$,
\begin{align}
\label{eq:FDT1}
\langle B(t) \rangle_{E} &=\; (\gamma e) E(t), \\ \label{eq:FDT2}
\langle B(t)B(s) \rangle_{E}
&= mk_B T\,\nu(t-s) + (\gamma e)^2 E(t)E(s),
\end{align}
Here, m is the effective mass of the system
coordinate, $k_B T$ is the bath temperature in energy units, $e$ the electrical charge, $\gamma \equiv g_E$ playing the role of an effective charge and $\nu(t - t')$ is the symmetrized noise kernel of the unbiased bath. The final and direct consequence of these chosen FDT relations in Eq.~(\ref{eq:FDT1})-(\ref{eq:FDT2}) is that the renormalization of the physically observable frequency \cite{Breuer2002} arises from the memory kernel where the bath specification is made \cite{boada2026field}. Coherent displacements of the pseudomode through a field bias treatment are therefore equivalent to linear displacement of the original mode(s) through the renormalization of the physically observable frequency \cite{boada2026field}.~Importantly, this allows us to incorporate intended mechanisms of control into the reduced model. 

As we'll discuss in the following sections, this distinction highlights nonlinearly intercoupled pseudomodes {as an avenue for \textit{equivalently} representing coherently driven interactions in the hardware by means of the noise model as opposed to time-dependence}: a coherent displacement can activate multi-photon parametric processes, confirming the faithfulness of the reduced model description.

\section{\\}\label{sec:general_theory}\vspace{-.3cm}
\subsection*{THEORETICAL METHODS:\\ Model Construction for Nonlinearly \\ Intercoupled Subsystem Ensembles}

While previous methods for the pseudomodal elimination of principal subsystem ensembles and nonlinear couplings have been considered \cite{Dalton2001,Dalton2003,GarrawayDalton2006, Pleasance2020,Pleasance2022, Menczel2024, Liang2024, Zhang2025}, the standardized extension into the theory for incorporating both elements and explicitly examining the repercussions has yet to be presented.

We undertake this task by constructing a continued-fraction Green's function from the retarded self-energy of the eliminated subsystem obtained through a Dyson expansion of the retarded Green's function~\cite{Dyson1949,FetterWalecka,Economou2006}.~Similar self-energy techniques have been widely employed to describe interacting many-body systems and the electromagnetic response of superconducting circuits~\cite{Beaudoin2011,MalekakhlaghPetrescuTureci2016,Keefe2025}, while Green's-function methods have previously provided the mathematical foundation underlying pseudomode representations of structured reservoirs~\cite{Garraway1997,Garraway1997b,Dalton2001,Pleasance2020}.~Rather than describing the eliminated subsystems with a single linear response function, however, we construct a family of (occupation-conditioned) Green's functions, each corresponding to a fixed configuration of the nonlinear retained subsystem~\footnote{~For any such configuration the eliminated subsystem reduces to an effective linear-response problem with shifted transition frequencies, allowing its Green's function to be reconstructed exactly from the corresponding self-energy.~The poles of each conditioned Green's function then define a corresponding pseudomode representation, while the collection of conditioned Green's functions provides a complete description of the nonlinear environment experienced by the retained subsystem.~In the linear limit the occupation dependence disappears and the formalism reduces identically to the standard construction}.

~Consequently, the formalism preserves the mathematical structure underlying conventional Green's-function pseudomode mappings while extending it to ensembles of nonlinearly intercoupled subsystems considered under the reduction, thereby explicitly extending pseudomode theory beyond linear reservoirs to nonlinear interacting subsystem ensembles of the kind considered here.

\subsection{\\
Subsystem-environment open \\ quantum system construction
}\label{sec: microscopic_treatment}

We consider a generic ensemble of nonlinearly intercoupled bosonic subsystems described by the Kerr Hamiltonian~\cite{Nigg2012}. We denote the complete collection of principal subsystems by $\{\mu_\mathrm{S}\}$, with $\mu,\nu\in\{\mu_\mathrm{S}\}$ labeling individual physical subsystems. Each subsystem $\mu$ is represented by the bosonic annihilation and creation operators $\hat a_\mu$ and $\hat a_\mu^\dagger$, with bare resonance frequency $\omega_\mu$. In a frame rotating at the applied drive frequencies, the effective detuning is given by $\omega_\mu-\Delta_\mu$, where $\Delta_\mu$ denotes the drive detuning from the bare resonance.~Local Kerr non-linearities are characterized by the anharmonicities $K_\mu$, while nonlinear intercouplings between distinct subsystems, $\mu$ and $\nu$, are described by the cross-Kerr strengths $\chi_{\mu\nu}$.~Throughout, $\alpha$ and $\beta$ index over the available transition channels.

Let $\hat H_{\rm sys}=\hat H_{\rm nd}+\hat H_{\rm d}$ denote the nonlinearly intercoupled subsystem Hamiltonian ensemble of the form
\begin{align}
\label{eq: kerrHamiltonian}
\hat H_{\rm nd}
&=\!\!\!\!
\sum_{\mu\in\{\mu_\mathrm{s}\}}
\!\!
\big[\,\,
(\omega_\mu-\Delta_\mu)\,\hat a_\mu^\dagger \hat a_\mu
-
%
\frac{K_\mu}{2}
(\hat a_\mu^\dagger \hat a_\mu)^2
\;\big]
\nonumber\\
&\quad
-\!\!\!\!
\sum_{\mu\in\{\mu_\mathrm{s}\}}
\!\!
\big[\,
\sum_{\nu\neq\mu}
\chi_{\mu\nu}
(\hat a_\mu^\dagger \hat a_\mu)
(\hat a_{\nu}^\dagger \hat a_{\nu})
\;\big]
,
\end{align}
while \vspace{-0.15cm}
\begin{equation}
\hat H_{\rm d}
=\!\!\!\!
\sum_{\mu\in\{\mu_\mathrm{d}\}}
\!\!
\big[\,
\!\!
\sum_{\nu\in\{\nu_\mathrm{d}\}}
(\omega_\nu^{(\mu)}-\Delta_\mu)\,\hat a_\nu^\dagger \hat a_\nu
\,\big],
\end{equation}
denotes the detuned drive modes while $\mu_\mathrm{d},\, \nu_\mathrm{d} \!\subset\!\{\mu_\mathrm{S}\}$ are the subset of displaced subsystems and total drive modes.~Importantly, Hamiltonians of this form \cite{Nigg2012} are generic to most superconducting circuit QED hardwares and therefore make Pleasance's framework for model reduction directly accessible to the language of superconducting processors.~As argued in \textbf{Appendix~\ref{app:nonlinear-green-function-derivation}.}, all the nonlinearly intercoupled subsystems of this form admit a rational polar decomposition in closed form, presented here in Section \textbf{\ref{sec:mode-elimination}}.~\ReviewerComment{A1}{Even if all of the subsystem-bath, drive-bath, and drive-subsystem couplings \cite{boada2026field} admit the simplistic bilinear form, 
\begin{equation}
\label{eq:bilinearcoupling}
\hat H_{\rm int}^{(\mu)}
=
\sum_{\mu}
\big[\,
\sum_{\alpha}
\sum_{\beta}
\left(
g_{\alpha\beta}^{(\mu)}\,
\hat a_\beta^\dagger \hat a_\alpha
+
g_{\alpha\beta}^{(\mu)*}\,
\hat a_\alpha^\dagger \hat a_\beta
\right)
\,\big],
\end{equation}
collected into the expression above through the global index~$\in\{\mu\}$, their explicit reduction becomes non-trivial because the drive displacement of the eliminated \textit{sector} modifies the channels seen by the retained mode}~\footnote{This scenario is intentional and commonly encountered in the dispersive readout of superconducting qubits, where linear displacements of the qubit are carried out indirectly.}.

\definecolor{ModeRed}{HTML}{992721}
\definecolor{ModeBlue}{HTML}{0D1582}
\definecolor{ModePurple}{HTML}{5E4877}
\definecolor{ModeOlive}{HTML}{8F865D}
\definecolor{ModeBlack}{HTML}{000000}
\begin{figure}
    \centering
    \begin{tikzpicture}[
        annot_label/.style={
            fill=white!85, 
            rounded corners, 
            draw=black!30, 
            inner sep=3pt, 
            font=\footnotesize
        }
    ]
    
    \node[anchor=south west, inner sep=0] (image) 
    {\includegraphics[width=0.95\linewidth]{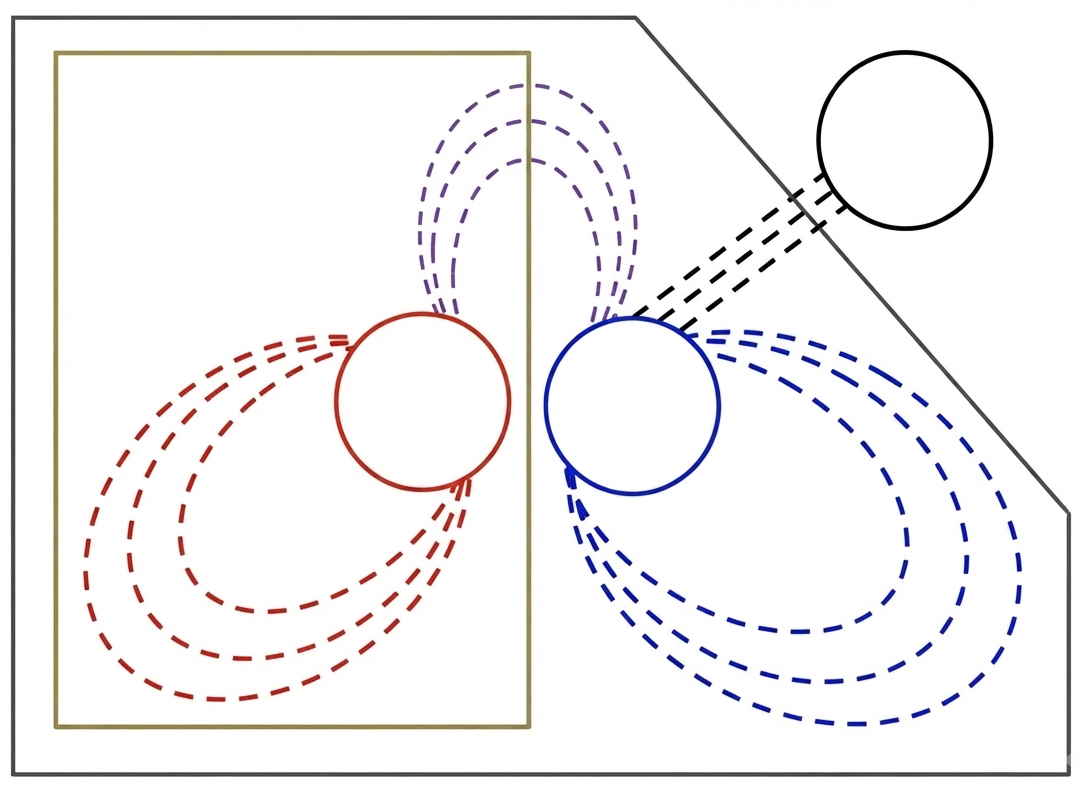}};
    
    
    \node[annot_label, text=ModeRed!70!black] (mode1) at (3.25, 2.9) {$\omega_a$};

    \node[annot_label, text=ModeBlue!70!black] (mode2) at (4.8, 2.9) {$\omega_b$};

    \node[annot_label] (mode3) at (6.85, 4.90) {$\omega_d$};

    \node[annot_label, text=ModePurple!70!black] (mode2) at (4.0, 4.0) {$\chi_{ab}$};

    \node[annot_label, text=ModeRed!70!black] (mode2) at (2.0, 2.0) {$\dfrac{K_a}{2}$};

    \node[annot_label, text=ModeBlue!70!black] (mode2) at (6.2, 2.0) {$\dfrac{K_b}{2}$};



    \end{tikzpicture}
    
    \caption{%
    A schematic of the hierarchy of eliminations considered here for the nonlinearly intercoupled principal mode ensemble.~The schematic depicts the complete (enlarged Markovian) Hilbert space of the three mode system with circles representing modes and dashed lines representing couplings.~The outer trapezoid represents the equivalent reduced model for the two modes and the inner green box represents the equivalent sequential reduced model which captures the influence of the displaced second mode over the remaining mode.%
    }
    \label{fig:hiearchy}
\end{figure} 

\Contribution{%
Hereafter, we use the term \enquote{sector} as opposed to subsystem, because the latter suggests a static partition of the Hilbert space.~A subsystem is a fixed physical entity; a \enquote{sector} is whatever collection of degrees of freedom is being traced over at any given stage of the reduction, and this collection is not fixed once and for all. As Fig.~\ref{fig:hiearchy} illustrates, the eliminated sector at one stage of the hierarchy — the outer trapezoid — may itself decompose into a retained piece and a further (\textit{sequentially}) eliminated sector at the next stage — the inner green box.~A sector thus references a role in the elimination hierarchy, assignable recursively to any grouping of modes.%
}

Let $\hat H_{\rm red}
$ denote the retained subsystem Hamiltonian, taken from the enlarged Hamiltonian $\hat H_{\rm sys}:=\hat H_{\rm nd}+\hat H_{\rm d}$, and let $\hat H_{\rm elim}
$ denote the eliminated subsystem Hamiltonian. Then, if $\hat H_{\rm elim}
$ admits a finite polar decomposition, $\exists\;\mathcal{L}_{\rm enl}\longrightarrow\mathcal{L}_{\rm red}$. We begin by partitioning the total subsystem Hamiltonian into retained and eliminated sectors,
$
\hat H_{\rm sys}
=
\hat H_{\rm red}
+
\hat H_{\rm elim}
+
\hat H_{\rm int}
$
where $\hat H_{\rm red}
$ may contain arbitrary nonlinear interactions among the retained subsystem operators, while $\hat H_{\rm elim}
$ describes the degrees of freedom to be traced out. 

The term $\hat H_{\rm int}
$ represents the interaction between retained and eliminated sectors. Let $\hat H_{\rm red}
$ and $\hat H_{\rm elim}
$ denote
\begin{equation}
\label{eq: reduced_Hamiltonian_definition}
\hat H_{\rm red}
=
\sum_{\mu
}
%
\big[\,\,
(\omega_\mu-\Delta_\mu)\,\hat a_\mu^\dagger \hat a_\mu
-
%
\frac{K_\mu}{2}
(\hat a_\mu^\dagger \hat a_\mu)^2
\;\big]
\end{equation}
and
\begin{align}
\hat H_{\rm elim}
&=
\sum_{\nu}
\big[\,
\epsilon_\nu
(t)\,\hat{a}_\nu^\dagger
+
\epsilon_\nu
(t)^{*}\,\hat{a}_\nu
\,\big].
\\
&\quad
+
\sum_{\nu 
}
%
\big[\,\,
(\omega_\nu-\Delta_\nu)\,\hat a_\nu^\dagger \hat a_\nu
-
%
\frac{K_\nu}{2}
(\hat a_\nu^\dagger \hat a_\nu)^2
\;\big]
\end{align}
while $\hat H_{\rm int}
$ is
\begin{equation}
\hat H_{\rm int}
=
\hat H_{\rm int}^{(\mu)}
-
\sum_{\mu
}
\sum_{\nu
}
\big[\, 
\chi_{\mu\nu}
(\hat a_\mu^\dagger\hat a_\mu)
(\hat a_\nu^\dagger\hat a_\nu)
\,\big].
\end{equation}

Here, $\epsilon_\nu(t)$ denotes the time-dependent coherent drive amplitude applied to subsystem $\nu$. Moving forward, we consider the reduction of the form $\in\{\mu\} \subsetneq \{\nu\}$
\begin{equation}
\label{eq: partitioned}
 H_{\rm tot}
=
 H_{\rm sys}
+
 H_{\rm bath}
+
 H_{\rm int},
\end{equation}
which leads to $\mathcal{L}_{\rm enl}$.~where $H_{\rm sys}$ is the subsystem Hamiltonian, while $ H_{\rm bath}$ denotes the Caldeira-Leggett bath.  

\subsection{\\ Heisenberg Equations of Motion \\ of the Enlarged Markovian Model
}

Working in the Heisenberg picture, the equation of motion for an arbitrary operator $\hat S_\alpha(t)$ acting on the retained subsystem follows from the partitioned Hamiltonian,
\begin{equation}
\label{eq:HeisenbergEOM}
    \frac{d}{dt}\hat S_\alpha(t)
    =
    -i[\hat S_\alpha(t),\hat H_{\mathrm{red}}^{(\mu)}]
    -i[\hat S_\alpha(t),\hat H_{\mathrm{int}}^{(\mu)}(t)],
\end{equation}
~For the coupling in Eq.~\eqref{eq:bilinearcoupling}, 
\begin{equation}
    \frac{d}{dt}\hat S_\alpha(t)
    =
    -i[\hat S_\alpha(t),\hat H_{\mathrm{red}}^{(\mu)}]
    +
    \hat F_\alpha^{(\mu)}(t),
\end{equation}
where
\begin{equation}
    \label{eq: totalforce}
    \hat F_\alpha^{(\mu)}(t)
    \equiv
    -i[\hat S_\alpha(t),\hat H_{\mathrm{int}}^{(\mu)}(t)]
\end{equation}
is the operator-valued force exerted by the eliminated sector.~For a retained operator $\hat S_\alpha(t)$, we write %
\begin{align} \label{eq:memorykernel}
\Dot{\hat S}_\alpha(t)
&=
-i[\hat S_\alpha(t),\hat H^{(\mu)}_\mathrm{red}] 
\\ \nonumber 
&+
\sum_{\beta}
\int_0^t d\tau\,
\mathcal{K}_{\alpha\beta}^{(\mu)}(t-\tau)\,
\hat S_\beta(\tau)
+
\hat \xi_\alpha^{(\mu)}(t). \nonumber
\end{align}
The general non-Markovian equation of motion were $\alpha$ and $\beta$ label the retained-sector coupling channels, while the superscript $(\mu)$ denotes that all modes except the retained
subsystem $(\nu \neq \mu)$ have been eliminated.~The memory kernel 
\begin{equation}
\mathcal{K}_{\alpha\beta}^{(\mu)}(t)
:=
-i\,\theta(t)\,\mathrm{Tr}
\langle
[
\hat F_\alpha^{(\mu)}(t),
\hat F_\beta^{(\mu)\dagger}(0)
]
\rangle_{\nu \neq \mu}.
\label{eq: kerneleq}
\end{equation}
is defined as the retarded response function of the eliminated sector projected onto the retained-sector coupling channels.~The expectation value $\mathrm{Tr}\langle \cdots \rangle_{\nu \neq \mu}$ denotes the partial trace over the eliminated sector, including both the eliminated subsystem modes and the residual Caldeira--Leggett bath. The force operator is
\begin{equation}
\hat F_\alpha^{(\mu)}(t)
=
\sum_{\beta\notin\mu}
[
g_{\alpha\beta}^{(\mu)}
\hat B_\beta^{(\mu)}(t)
+
g_{\alpha\beta}^{(\mu)*}
\hat B_\beta^{(\mu)\dagger}(t)
],
\end{equation}
where $\beta \notin\{\mu\}$ indexes over eliminated degrees of freedom and $ B_\beta
(t)$ denotes the corresponding eliminated-sector
operator evolved under $H_{\rm elim}
$. 

The inhomogeneous term, $\hat \xi_\alpha
(t)$, is the noise operator associated with the initial state of the eliminated sector.~For a single uncoupled eliminated oscillator, $G_\beta
(t)=\mathrm{exp}\{-i\omega_{\alpha\beta} t\}$, while in the interacting case $G_\beta
(t)$ is replaced by the appropriate channel-resolved Green's function of the eliminated sector, 
\begin{equation}
\hat \xi_\alpha^{(\mu)}(t)
=
\sum_{\beta\notin\mu}
[
g_{\alpha\beta}^{(\mu)}
\hat B_\beta^{(\mu)}(0)\,
G_\beta^{(\mu)}(t).
+
\text{h.c.}
]
\end{equation}
Passing over to the frequency domain through an inverse Laplace transform defines the matrix-valued self-energy,
\begin{equation}
\label{eq:selfenergyKernel}
\Sigma_{\alpha\beta}^{(\mu)}(\omega)
=
\int_0^\infty dt\,
e^{i\omega t}
\mathcal{K}_{\alpha\beta}^{(\mu)}(t).
\end{equation}
Equivalently, $\Sigma_{\alpha\beta}^{(\mu)}(\omega)$ is the projected susceptibility of the eliminated sector,
\begin{align}
\Sigma^{(\mu)}(\omega)&=
\sum_{\alpha\in\mu}
\sum_{\beta\in\nu}
g_{\alpha\beta}^{(\mu)}
\,
\widetilde G_{\alpha\beta}^{(\mu)}(\omega)
\end{align}%
where $\widetilde G_{\alpha\beta}^{(\mu)}(\omega)$ denotes the Laplace transform of the eliminated sector's ($\nu$) retarded Green's function (see also: \textbf{Appendix \ref{app: closure}}).~The retained sector ($\mu$) Green's function therefore obeys the Dyson equation
\begin{equation}
\label{eq:Dyson_equation}
[
\widetilde G^{(\mu)}(\omega)
]^{-1}
=
[
\widetilde G_0^{(\mu)}(\omega)
]^{-1}
-
\Sigma^{(\mu)}(\omega).
\end{equation}
Here $\widetilde G_0^{(\mu)}(\omega)$ is the Green's function generated by the isolated retained Hamiltonian, $\hat H^{(\mu)}$, while $\Sigma^{(\mu)}(\omega)$ contains the full influence of $H_\mathrm{tot}=\hat H_{\rm elim}^{(\mu)}+\hat H_{\rm int}^{(\mu)}$.~When $\Sigma^{(\mu)}(\omega)$ is rational, its poles determine the finite pseudomode representation; otherwise, the pseudomode model can be obtained by a controlled rational approximation of the retarded self-energy, $\Sigma^{(\mu)}(\omega)$.

The (retarded) self-energy defined in the Dyson-type equation for the retained subsystem is obtained by projecting the eliminated subsystem onto a fixed occupation of the retained subsystem.~If the eliminated subsystem possesses discrete dressed eigenstates, the resulting occupation-conditioned self-energy admits the polar decomposition of the continued fraction
\begin{equation}
\Sigma^{(\mu)}(\omega)
=
\sum_{\alpha}\sum_{\beta}
\frac{
|\Lambda_{\alpha\beta}^{(\mu)}|^2
}
{
\omega-\Omega_{\alpha\beta}^{(\mu)}
+i\kappa_{\alpha\beta}^{(\mu)}/2
},
\label{eq:SigmaPole}
\end{equation}
The quantities $\Omega_{\alpha\beta}^{(\mu)}$ and $\kappa_{\alpha\beta}^{(\mu)}$ and $|\Lambda_{\alpha\beta}^{(\mu)}|^2$ denote the corresponding occupation-conditioned dressed transition frequencies, linewidths, and denote the coupling weight of each pole to the retained subsystem where $\alpha$ labels the allowed transition channels of the eliminated subsystem and $\beta$ labels the poles associated with each channel.


\begin{figure}[t]
    \centering
    \begin{tikzpicture}[
        annot_label/.style={
            fill=white!80, 
            rounded corners, 
            draw=black!30, 
            inner sep=3pt, 
            font=\footnotesize
        }
    ]
    
    \node[anchor=south west, inner sep=0] (image) 
    {\includegraphics[width=0.8\linewidth]{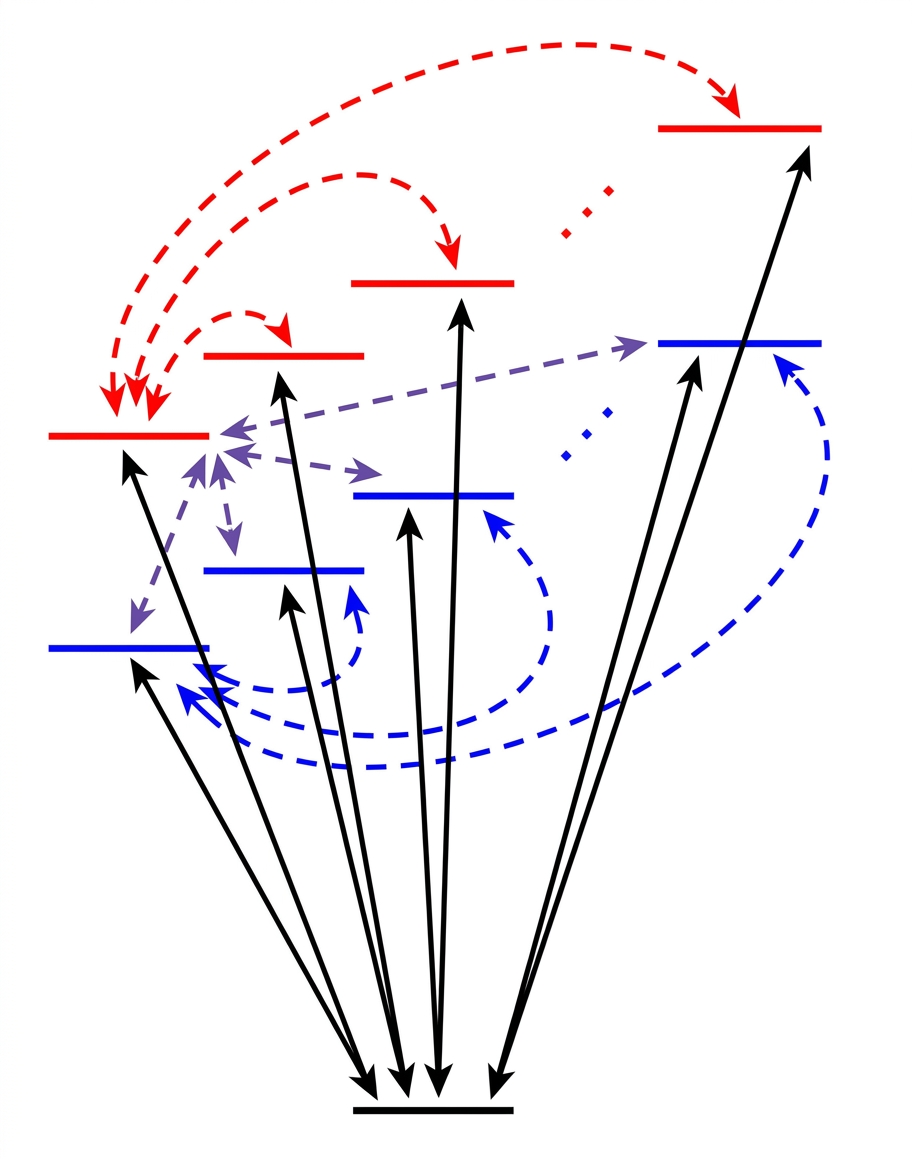}};
    
    
    \node[annot_label] (ground_label) at (4.5, 0.5) {$|0\rangle_{d}$};
    
    
    \node[annot_label, text=blue!70!black] (mode_m_label) at (6.9, 6.25) {$|m\rangle_{b}$};
    \node[annot_label, text=red!70!black] (mode_m_label) at (6.9, 8.0) {$|n\rangle_{a}$};

    

    \end{tikzpicture}
    
    \caption{Energy level diagram for the nonlinearly intercoupled two-mode prototype. Ground state of the auxiliary system is coupled to upper states 1,2,3, \dots of two distinct modes $\{m, n\}$ by the transitions with frequencies $\omega_1, \omega_2, \omega_3, \dots$. The red and blue dashed lines represent self-Kerr interactions for each mode, while the purple dashed lines represent cross-Kerr coupling between the upper states.}
    \label{fig:placeholder}
\end{figure}

\subsection{Pseudomode Construction}\label{sec: pseudomode_construction}

Before carrying out the rational polar decomposition, the retained subsystem self-energy may be written as the sum of channel-resolved contributions,
\begin{align}
\Sigma^{(\mu)}(\omega)
&=
\sum_{\alpha}
\Sigma_{\alpha}^{(\mu)}(\omega)
\\&=
\sum_{\alpha}
\big( 
\sum_{\beta}
\frac{
|\tilde{\Gamma}_{\alpha}^{(\mu)}|^2
\gamma_{\alpha\beta}^{(\mu)}
}
{
\omega-z_{\alpha\beta}^{(\mu)}
}
\big)
\end{align}
where each contribution is obtained from the (dressed) Green's function of the corresponding eliminated-sector where $|\tilde{\Gamma}_{\alpha}^{(\mu)}|^2$ denotes the coupling strength between the retained subsystem and eliminated transition channel $\alpha$,
\begin{equation}
\Sigma_{\alpha}^{(\mu)}(\omega)
=
|\tilde{\Gamma}_{\alpha}^{(\mu)}|^2
G_{\alpha}^{(\mu)}(\omega).
\label{eq:channelSigma}
\end{equation}
 Consequently,
\begin{equation}
G_{\alpha}^{(\mu)}(\omega)^{-1}
=
\omega-\Omega_{\alpha}^{(\mu)}
+i\kappa_{\alpha}^{(\mu)}/2
-
\Pi_{\alpha}^{(\mu)}(\omega)
,
\label{eq:Greendressed}
\end{equation}
where $\Pi_{\alpha}^{(\mu)}(\omega)$ denotes the self-energy acquired by the eliminated transition due to its coupling to the rest of the eliminated subsystem.~
Whenever this self-energy is rational, it also admits a finite polal decomposition,
\begin{equation}
\Pi_{\alpha}^{(\mu)}(\omega)
=
\sum_{\beta}
\frac{
|\eta_{\alpha\beta}^{(\mu)}|^2
}
{
\omega-\omega_{\alpha\beta}^{(\mu)}
},
\label{eq:PiPole}
\end{equation}
where $\omega_{\alpha\beta}$ are the poles generated by the remaining eliminated degrees of freedom.~Substituting Eq.~(\ref{eq:PiPole}) into Eq.~(\ref{eq:Greendressed}) produces a rational Green's function whose poles are determined by the zeros of the dressed denominator,
$
\omega
-
\Omega_{\alpha}^{(\mu)}
+
i\kappa_{\alpha}^{(\mu)}/2
-
\Pi_{\alpha}^{(\mu)}(\omega)
=
0.
$ Performing the partial-fraction decomposition gives
\begin{equation}
G_{\alpha}^{(\mu)}(\omega)
=
\sum_{\beta}
\frac{
\gamma_{\alpha\beta}^{(\mu)}
}
{
\omega-z_{\alpha\beta}^{(\mu)}
},
\label{eq: decomposition}
\end{equation}
where $\gamma_{\alpha\beta}^{(\mu)}$ are the residues of the dressed poles.~Substituting this result back into Eq.~(\ref{eq:channelSigma}) and identifying
$
|\Lambda_{\alpha\beta}^{(\mu)}|^2
=
|\tilde{\Gamma}_{\alpha}^{(\mu)}|^2
\gamma_{\alpha\beta}^{(\mu)}
$
recovers the occupation-conditioned pole expansion of Eq.~(\ref{eq:SigmaPole}).~These poles can be obtained by diagonalizing the effective non-Hermitian generator of the eliminated transition channel,
$
\mathcal H_{\alpha}
=
H_{\alpha}
-
i/2\mathcal{B}_{\alpha},
$
where $H_{\alpha}$ contains the coherent dynamics within transition channel $\alpha$, whereas $\mathcal{B}_{\alpha}$ describes the corresponding irreversible self-dressing processes.~Solving
$
\mathcal H_{\alpha}
|\phi_{\alpha\beta}\rangle
=
z_{\alpha\beta}^{(\mu)}
|\phi_{\alpha\beta}\rangle
$ 
yields the complex dressed poles
\begin{equation}
z_{\alpha\beta}^{(\mu)}
=
\Omega_{\alpha\beta}^{(\mu)}
-
i
\kappa_{\alpha\beta}^{(\mu)}/2,
\end{equation}
whose real and imaginary parts define the occupation-conditioned transition frequencies and couplings,
\begin{equation}
\label{eq: poles_solution}
\Omega_{\alpha\beta}^{(\mu)}
=
\operatorname{Re}
[
z_{\alpha\beta}^{(\mu)}
],
\quad 
\kappa_{\alpha\beta}^{(\mu)}
=
-2
\operatorname{Im}
[
z_{\alpha\beta}^{(\mu)}
].
\end{equation}

The corresponding coupling weight, $|\Lambda_{\alpha\beta}^{(\mu)}|^2$, is determined by the overlap of the retained-subsystem coupling operator with the dressed eigenstates of the eliminated sector. Consequently, the spectral density observed by the retained subsystem is the double-sum over $\alpha,\,\beta$ in
\begin{equation}
J^{(\mu)}_{\alpha\beta}(\omega)
=
\frac{
|\Lambda_{\alpha\beta}^{(\mu)}|^2
}{\pi}
\frac{
\kappa_{\alpha\beta}^{(\mu)}/2
}{
\big[
\omega-\Omega_{\alpha\beta}^{(\mu)}
\big]^2
+
\big[
\kappa_{\alpha\beta}^{(\mu)}/2
\big]^2
}.
\label{eq:spectraldensitypseudo}
\end{equation}

The poles, $z_{\alpha\beta}^{(\mu)}$, are therefore obtained by diagonalizing the eliminated-sector Hamiltonian at fixed occupation of the retained subsystem, providing the most direct route to the spectral density entering the reduced non-Markovian description. 

\subsection{
\\Generalized Auxiliary \& Resultant \\ Reduced Non-Markovian Channel Models
}\label{sec: auxiliary_equation}

Consider an arbitrary collection of subsystem transition operators %
$
\mathbf{\mathcal{L}}
=
\{
\mathcal L_1, \mathcal L_2,\ldots
\},
$ %
where each operator $\mathcal L_j$ labels an independent dissipative channel.~Typical examples for $\mathcal L$ could include terms such as $\mathcal L_j\in \{
a_\mu,\;
a_\mu^2,\;
a_\mu^\dagger a_\mu,\;
a_\mu^\dagger a_\nu,\;
a_\mu^\dagger a_\nu a_\nu, \ldots 
\},$ although no specific restriction has been imposed on their algebraic form.~The enlarged Hamiltonian assumes the universal form
\begin{align}
\label{eq: aux1}
H_{\rm aux}
&=
H^{(\mu)}_\mathrm{red}
+
\sum_j
H_{\rm PM}^{(j)}
+
\sum_j
H_{\rm int}^{(j)},
\end{align}
where
\begin{align}
\label{eq: aux2}
H_{\rm PM}^{(j)}
&=
\sum_\ell
\Omega_\ell^{(j)}
b_\ell^{(j)\dagger}
b_\ell^{(j)}
+
H_{\rm PM,nl}^{(j)},
\\
\label{eq: aux3}
H_{\rm int}^{(j)}
&=
\sum_\ell
\left(
\Gamma_\ell^{(j)}
L_j^\dagger
b_\ell^{(j)}
+
{\rm h.c.}
\right)
+
H_{\rm int,nl}^{(j)}.
\end{align}
For each transition channel $\mathcal L_j$, the enlarged Markovian model
introduces an auxiliary pseudomode $
b_j
$ whose dynamics reproduce the projected Green's function of the eliminated sector associated with that channel.~Here the superscript $(j)$ labels the subsystem transition channel, while the index $\ell$ labels the poles required to represent the projected Green's function of that channel.~The terms $H_{\rm PM,nl}$ and $H_{\rm int,nl}$ collect arbitrary nonlinear interactions inherited from the eliminated subsystem.

~Eliminating the auxiliary modes yields the channel-resolved self-energy from which the corresponding spectral density follows.
\begin{equation}
\Sigma_j(\omega)
=
\sum_\ell
\frac{|\Gamma_\ell^{(j)}|^2}
{\omega-\Omega_\ell^{(j)}
+i\kappa_\ell^{(j)}/2}.
\end{equation}
The combined environmental response is therefore the sum over all
transition channels, $\in J_{j}(\omega)$, 
\begin{equation}
J_j(\omega)
=
\sum_\ell
\frac{|\Gamma_\ell^{(j)}|^2
\kappa_\ell^{(j)}/2}
{
(\omega-\Omega_\ell^{(j)})^2
+
(\kappa_\ell^{(j)}/2)^2
}.
\end{equation}
Each channel contributes an independent non-Markovian decay kernel,
\begin{equation}
\kappa_j(t)
=
2
\sum_\ell
|\Gamma_\ell^{(j)}|^2
{\rm Re}
\Big[
\frac{
1-
e^{-[\kappa_\ell^{(j)}/2+i\Omega_\ell^{(j)}]t}
}
{
\kappa_\ell^{(j)}/2
+
i\Omega_\ell^{(j)}
}
\Big],
\end{equation}
which approaches the Markovian limit as $t\rightarrow\infty$.~The corresponding reduced quantum master equation therefore assumes the
universal form
\begin{equation}
\label{eq: mastereqgeneral}
\dot\rho_\mu
=
-i
[
H^{(\mu)}_\mathrm{red}
+
H_{\rm LS}(t),
\rho_\mu
]
+
\sum_j
\frac{\kappa_j(t)}{2}
\mathcal D[L_j]\rho_\mu,
\end{equation}%
where every subsystem transition is associated with its own projected Green's function which leads to the spectral density that enters auxiliary pseudomode representation that determine the non-Markovian memory kernels.\newpage 

\Contribution{%
One advantage of the specified nonlinearity in Eq.~\eqref{eq: kerrHamiltonian} is that every model considered here is an instance of the two-mode prototype illustrated in Fig.~\ref{fig:hiearchy} and presented in Sec.\textbf{~\ref{sec:two_mode}.}: each retained mode obeys a master equation of the form (see also: \textbf{Appendix \ref{app:redfield}.-\ref{app:sequentialelims}.})
\begin{equation}
\dot{\rho}_\mu = -\frac{i}{\hbar}\big[\hat{H}_{\rm red}^{\rm LS},
\rho_\mu\big] + \frac{\kappa_1(t)}{2}\,\mathcal{D}[\hat{a}]\rho_\mu
+ \frac{\kappa_2(t)}{2}\,\mathcal{D}[\hat{a}^2]\rho_\mu,
\label{eq:master_prime}
\end{equation}
with $\mathcal{D}[\hat{L}]\rho \equiv \hat{L}\rho\hat{L}^\dagger - \{\hat{L}^\dagger\hat{L},\rho\}/2$.~where $\kappa_1(t)$ is the rate associated with the single-exchange channel, while $\kappa_2(t)$ is the rate of a higher-order channel activated through a coherent displacement of an eliminated mode in Sec~\textbf{\ref{sec:coherentdisplacement}}. 
}

\section{\\}\vspace{-0.25cm}\label{sec:mode-elimination}
\subsection*{THEORETICAL RESULTS I:\\ Select Reductions \& Closed-form \\ Green's Function Solutions}
We now apply the general framework from Section\textbf{~\ref{sec:general_theory}.} to select systems.\footnote{For the following sections, we will denote the bosonic operators $a_{(i)} (a_{(i)}^\dagger) = a, (a^\dagger)$ with $i = \mu,\nu$.} We begin with the simplest case: two bosonic modes $a$ and $b$, each with self-Kerr, coupled through a cross-Kerr term and a bilinear exchange interaction,
$H = H_0 + V_{0}$ with
\begin{align}
H_0 &= \omega_a a^\dagger a - \frac{K_a}{2}a^{\dagger 2}a^2 \\&
+ \omega_b b^\dagger b - \frac{K_b}{2}b^{\dagger 2}b^2 
- \chi_{ab}\, a^\dagger a\, b^\dagger b, \nonumber
\end{align}
and bilinear coupings
\begin{equation}
V_0 = g_{ab}\!\left(a^\dagger b + b^\dagger a\right).
\end{equation}
The total excitation number $N = a^\dagger a + b^\dagger b$ is conserved, $[H,N]=0$, so the Hilbert space decomposes into invariant total-excitation sectors.~The two-mode prototype is treated in full to establish the master template; the three- and four-mode cases follow by a direct substitution. 

\subsection{The two-mode prototype}\label{sec:two_mode}

In this subsection, we'll consider the explicit elimination of the form
\begin{align}
    H_\mathrm{keep} &= \omega_a \, a^\dagger a - \frac{K_a}{2}a^{\dagger 2}a^2\\ 
    H_\mathrm{elim} &= \omega_b \, b^\dagger b - \frac{K_b}{2}b^{\dagger 2}b^2 \\
    H_\mathrm{int} &= -\chi_{ab}\, a^\dagger a\, b^\dagger b + g_{ab}\!\left(a^\dagger b + b^\dagger a\right)
\end{align}

Starting from the bare Fock basis $|n,m\rangle = |n\rangle_a \otimes |m\rangle_b$ with energies denoted as
\begin{align}
\label{eq: energy_twomode}
E_{n,m}
&=
n\omega_a+m\omega_b
\\&\nonumber
-\frac{K_a}{2}n(n-1)
-\frac{K_b}{2}m(m-1)
-\chi_{ab}nm.
\label{eq:energy_two_mode}
\end{align}
the exchange term connects
$|n,m\rangle \leftrightarrow |n-1,m+1\rangle$
by the matrix element
$
\langle n-1,m+1|\,H_{\mathrm{int}}\,|n,m\rangle
=
g_{ab}\sqrt{n(m+1)}
$.\newpage

~The local transition frequencies of the two participating modes are therefore
\begin{align}
\Omega^{\,(a)}_{(n-1,\,m)} &= \omega_a - K_a(n-1) - \chi_{ab}\, (m), \\
\Omega^{\,(b)}_{(n-1,\,m)} &= \omega_b - K_b\, (m) - \chi_{ab}(n-1),
\end{align}
where the self-Kerr, $K_i$, of each mode shifts according to its own occupation, while the cross-Kerr, $\chi_{ij}$, shifts each line  according to the occupation of the other.

The reduced dynamics of the retained subsystem amplitudes are obtained after eliminating the environmental (and pseudomodal) degrees of freedom which then lead to the integro-differential equation
\begin{equation}
\dot{c}_{n,m}(t)
=
-\int_0^t dt'\,
K_{n,m}(t-t')\,
\widetilde c_{n,m}(t'),
\end{equation}
where $\mathcal{K}_{n,m}(t - t')$ is the memory kernel arising from the coupling to the eliminated sector~\cite{Breuer2002}.~Passing over to the frequency domain,
\begin{gather}
    \tilde{c}_{n,m}(z)
    =
    \int_0^\infty dt \, e^{i z t} \, c_{n,m}(t)
    \\
    c_{n,m}(t)
    =
    \int_{-\infty + i\epsilon}^{\infty + i\epsilon}
    \frac{d z}{2\pi} \,
    e^{-i z t} \, \tilde{c}_{n,m}(z)
\end{gather}
the convolution becomes algebraic and the solution of the retained channel can be written explicitly in terms of the Green's function where $\Delta\Omega_{n-1,m}$ denotes the local occupation-conditioned transition frequency of the retained mode.~For a single exchange channel, %
\begin{align}
    \Sigma_{n,m}(\omega)
    = \frac{|\mathcal{M}_{n,m}|^2}{\omega - E_{n-1,m+1}}
\end{align}
where $E_{n-1,m+1}$ is the energy of the intermediate state in which one excitation has been transferred to the $k$th environmental mode through exchange channel $k$, while
\begin{equation}
\mathcal{M}_{n,m}
=
\left\langle
n-1,m+1
\right|
H_{\mathrm{int}}^{(k)}
\left|
n,m
\right\rangle
\end{equation}
leads to the channel coupling strength and the pseudomodal elimination for $H_\mathrm{keep}$ and follows naturally from the original treatment as summarized below.\newline 

\paragraph{Exact local elimination.} Projecting onto an arbitrary state of the enlarged system $\Psi(t)$,
\begin{align}
&|\Psi(t)\rangle \\&= c_{n,m}(t)|n,m\rangle 
+ \exchangeterm c_{n-1,m+1}(t)|n-1,m+1\rangle. \nonumber 
\end{align} 
The Schr\"odinger equation gives
\begin{align}
i\dot c_{n,m}(t) &= E_{n,m}c_{n,m}(t) 
+ \exchangeterm
c_{n-1,m+1}(t),\\ \nonumber 
i\dot c_{n-1,m+1}(t) &= \\&\!\!\!\!\! E_{n-1,m+1}c_{n-1,m+1}(t) 
+ 
\exchangeterm
c_{n,m}(t).
\end{align}

In the chosen frame, after time $t$, $c_{n-1,m+1}(0)$ evolves to $c_{n-1,m+1}(t)$. \newpage

Eliminating $c_{n-1,m+1}(t)$ exactly,
\begin{align}
&c_{n-1,m+1}(t) =
\\&\nonumber 
-i
\exchangeterm
\int_0^t\! dt'\, 
e^{-iE_{n-1,m+1}(t-t')}c_{n,m}(t'),
\end{align}
and substituting into the first equation and factoring out the bare phase $c_{n,m}(t) = e^{-iE_{n,m}t}\tilde c_{n,m}(t)$ yields
\begin{align}\label{eq:nonlocal_eom_2mode}
&\dot{\tilde c}_{n,m}(t) =\\\nonumber&\quad 
-(\exchangeterm)^2\!\int_0^t\! dt'\,
e^{i(E_{n,m}-E_{n-1,m+1})(t-t')}\tilde c_{n,m}(t')
\end{align}
after factoring out $c_{n,m}(t) = e^{-iE_{n,m}t}\tilde c_{n,m}(t)$.~Recalling our definition of $\exchangeterm$ leads to $\exchangeterm^2= g_{ab}^2 n(m+1).$ The phase difference (meaning the relative phase accrued between $c(0)$ and $c(t)$) reduces via 
\begin{align}&E_{n,m} - E_{n-1,m+1} =\\ &\quad\quad\Omega^{(a)}_{(n-1,m)} - \Omega^{(b)}_{(n-1,m)}=\Delta\Omega^{(ab)}_{(n-1,m)}\end{align} into the local two-level exchange problem with (energy-dependent) bare frequencies and coupling rates.\newline 

\paragraph{Green's functions and self-energies.}
In frequency space the retained-channel Green's function is
\begin{equation}\label{eq:green_2mode}
G_{(n,m)}^{(a)}(z)
=
\frac{1}
{z-\Omega_{(n-1,m)}^{(a)}-\Sigma_{(n,m)}^{(a)}(z)},
\end{equation}
with self-energy
\begin{equation}\label{eq:selfenergy_2mode}
\Sigma_{(n,m)}^{(a)}(z)
=
\frac{g_{ab}^{\,2}\, n(m+1)}
{z-\Omega_{n-1,m}^{(b)}}.
\end{equation}

Here, $\Omega_{(n-1,m)}^{(a)}$ and $\Omega_{(n-1,m)}^{(b)}$ are the local occupation-conditioned transition frequencies of the retained and eliminated modes, respectively.~The dressed poles are the roots of
\begin{equation}
\big[z-\Omega_{n-1,m}^{(a)}\big]
\big[z-\Omega_{n-1,m}^{(b)}\big]
-
g_{ab}^{\,2}n(m+1)
=0,
\end{equation}
and therefore with $\Delta\Omega^{(ab)}_{(n-1,m)}\!\!\!
=
\Omega_{(n-1,m)}^{(a)}
-
\Omega_{(n-1,m)}^{(b)},$
\begin{gather}\label{eq:poles_2mode}
z_{n,m,\pm}^{(a)}
=
\frac{
\Omega_{(n-1,m)}^{(a)}
+
\Omega_{(n-1,m)}^{(b)}
}{2}
\pm \quad\quad\quad\quad\quad\quad
\\
\quad\quad\frac{1}{2}
\sqrt{
(\Delta\Omega^{(ab)}_{(n-1,m)})\,\!^2
+
4g_{ab}^{\,2}n(m+1)
},
\end{gather}
Notably, what we have done is define the local frequencies associated with the interaction, and the pseudomodal elimination follows naturally from the original treatment.
\newline  

\paragraph{Fixed-sector in tri-diagonal form.}
Within the fixed total-excitation sector $N=n+m$, the basis
$|r;N\rangle\equiv |r,N-r\rangle$ for $r=0,1,\dots,N$ renders
the Hamiltonian tridiagonal, which has a number of desirable properties.

\newpage 

The entries are
\begin{multline}
H^{(N)} = \sum_{r=0}^{N} E_r^{(N)}|r;N\rangle\langle r;N| \\
+ \sum_{r=0}^{N-1} J_{r+1}^{(N)}\!\left(|r{+}1;N\rangle\langle r;N| 
+ \mathrm{h.c.}\right),
\end{multline}
with $E_r^{(N)}$ being
\begin{align}
E_r^{(N)}
&=
r\omega_a+(N-r)\omega_b \nonumber
\\&-\frac{K_a}{2}r(r-1)
-\frac{K_b}{2}(N-r)(N-r-1)
\\&-\chi_{ab}r(N-r), \nonumber
\end{align}
with {jump} amplitude $J_{r+1}^{(N)} = \tilde{g}\sqrt{(r+1)(N-r)}$, where $\tilde{g}:=g_{ab}\in\mathbb{R}$ for our two-mode prototype. The projected resolvent on $|r;N\rangle$ admits an exact continued-fraction self-energy,
\begin{equation}\label{eq:cf_general}
\Sigma_r^{(N)}(z) = \Sigma_{r,-}^{(N)}(z) + \Sigma_{r,+}^{(N)}(z),
\end{equation}
with
\begin{equation}
\Sigma_{r,\pm}^{(N)}(z) = \frac{(J_{r\pm 1}^{(N)})^2}
{z - E_{r\pm 1}^{(N)} - \Sigma_{r\pm 1,\pm}^{(N)}(z)},
\end{equation}
with the boundary conditions $\Sigma_{0,-}^{(N)} = \Sigma_{N,+}^{(N)} =0$.~Thus, equations ~\eqref{eq:green_2mode}-\eqref{eq:cf_general} constitute the master template 
referenced by the multi-mode cases below:~the eliminated mode contributes a single, occupation-conditioned pole to the self-energy, the dressed poles are the roots of a quadratic, and 
the fixed-sector global form is tridiagonal with a continued-fraction self-energy.

\begin{table*}[t]
\small  
\centering
\renewcommand{\arraystretch}{1.7}
\begin{tabular}{llll}
\hline\hline
&
Two-mode (bilinear) $\qquad$
&
Three-mode (3-wave) $\qquad\qquad$
&
Four-mode (bilinear)
\\
\hline
Coupling
&
$g_{ab}(a^\dagger b+\mathrm{h.c.})$
&
$g_{3}(abc^\dagger+\mathrm{h.c.})$
&
$g_{4}(cd^\dagger+\mathrm{h.c.})$
\\
Channel
&
$|n,m\rangle \leftrightarrow |n{-}1,m{+}1\rangle$
&
$|n,m,\ell\rangle \leftrightarrow |n{-}1,m{-}1,\ell{+}1\rangle$
&
$|n,m,\ell,k\rangle \leftrightarrow |n,m,\ell{-}1,k{+}1\rangle$
\\
Exchange $|\mathcal{M}|^2$
&
$n(m+1)$
&
$nm(\ell+1)$
&
$\ell(k+1)$
\\
$\Omega_\alpha$
&
$\omega_a-K_a(n-1)-\chi_{ab}m$
&
$\omega_a+\omega_b
-K_a(n-1)
-K_b(m-1)$
&
$\omega_c-K_c(\ell-1)$
\\
&
&
$\qquad
-\chi_{ab}(n+m-1)
-(\chi_{ac}+\chi_{bc})\ell$
&
$\qquad
-\chi_{ac}n
-\chi_{bc}m$
\\
$\Omega_\beta$
&
$\omega_b-K_bm-\chi_{ab}(n-1)
\quad\quad
$
&
$\omega_c-K_c\ell
-\chi_{ac}(n-1)
-\chi_{bc}(m-1)
\quad\quad$
&
$\omega_d$
\\
Conserved
&
$N=n+m$
&
$Q=n+m+2\ell,\qquad D=n-m$
&
$n,\,m^\ast;\qquad N=n+m+\ell+k$
\\
Tridiagonal $J_{r+1}$
&
$g_{ab}\sqrt{(r+1)(N-r)}$
&
$g_{3}\sqrt{(A-r)(B-r)(r+1)}$
&
$g_{4}\sqrt{(N-A-B-r)(r+1)}$
\\
\hline\hline
\end{tabular}
\caption{
Substitutions into the master template
Eqs.~\eqref{eq:green_2mode}--\eqref{eq:poles_2mode}
for the two-, three-, and four-mode reductions.
The two-mode prototype provides the universal local Green's-function template; the higher-order cases follow by replacing the occupation-conditioned transition frequencies and exchange matrix element, spectator modes enter only via Kerr shifts of $\Omega_\alpha$ and $\Omega_\beta$.~The $^\ast$ is used to indicate that $n$ and $m$ are each separately conserved (rather than only a combination of them).
}
\label{tab:cases}
\end{table*}

\subsection{\\
Multi-Mode \& Higher-Order Coupling \\ Generalizations for Subsystem Ensembles
}\label{sec:multimodes}

The construction presented above extends directly to three- and four-mode 
systems with three-wave-mixing or bilinear exchange couplings.~The Schr\"odinger equation, Laplace transform, and phase-factoring steps in the previous section proceed identically; only the conserved quantities, the matrix elements, and the occupation-conditioned frequencies change.~We collect the substitutions for a selection of eliminations in Table~I.\newline 

\paragraph{
Three-mode undergoing a mixing scheme.}%
Three bosonic modes $a, b, c$ with Kerr-plus-cross-Kerr 
$H_0$ and three-wave-mixing coupling
\begin{equation}
\label{eq: mixing scheme coupling}
V_3 = g_3\left(a b c^\dagger + a^\dagger b^\dagger c\right)
\end{equation}
support the local exchange in the channel $|n,m,\ell\rangle \leftrightarrow |n{-}1,m{-}1,\ell{+}1\rangle$.~We retain the Kerr modes $a$ and $b$ and eliminate mode $c$:
\begin{align}
  H_{\mathrm{keep}} &= \omega_a a^\dagger a 
    + \omega_b b^\dagger b 
    - \frac{K_b}{2}\,b^{\dagger 2}b^2
    - \frac{K_a}{2}\,a^{\dagger 2}a^2
    \nonumber\\
    &\quad - \chi_{ab}\, a^\dagger a\, b^\dagger b,
  \label{eq:Hkeep3}\\
  H_{\mathrm{elim}} &= \omega_c c^\dagger c - \frac{K_c}{2}\,c^{\dagger 2}c^2,
  \label{eq:Helim3}\\
  H_{\mathrm{int}} &= V_3
    - \chi_{ac}\, a^\dagger a\, c^\dagger c
    - \chi_{bc}\, b^\dagger b\, c^\dagger c.
  \label{eq:Hint3}
\end{align}
The interaction thus conserves $Q = a^\dagger a + b^\dagger b + 2c^\dagger c$ and $D = a^\dagger a - b^\dagger b$. Substituting the entries from the second column of Table I.
yields the dressed poles. Within fixed $(Q,D)$ sectors, the basis $|r;Q,D\rangle = |A{-}r, B{-}r, r\rangle$ with $A=(Q+D)/2$,~$B=(Q-D)/2$ therefore renders the Hamiltonian tridiagonal with jump term \begin{equation}\label{eq: jumpterm3}
J_{r+1}^{(Q,D)} = g_{3}\sqrt{(A-r)(B-r)(r+1)}\end{equation}%
and Eq.~\eqref{eq:cf_general} applies verbatim.~Applying the elimination procedure and denoting
\begin{equation}
\label{eq: G3_greens}
  G^{(ab)}_{n,m,\ell}(z) \longrightarrow G_3(z) =
  \frac{1}{z - \Omega^{(ab)}_{(n-1,m-1,\ell)} - \Sigma_3(z)}
\end{equation}
\begin{equation}
  \Sigma^{(ab)}_{n,m,\ell}(z) \longrightarrow \Sigma_3(z) =
  \frac{|\mathcal{M}_3|^2}{z - \Omega^{(c)}_{(n-1,m-1,\ell)}},
  \label{eq:G3Sigma3}
\end{equation}
with dressed poles following from Eq.~\eqref{eq:Greendressed} by direct substitution.~The mixing term therefore couples the composite transition in $(a,b)$ to a single transition in mode $(c)$ through the matrix element~$|\mathcal{M}|^2 \longrightarrow|\mathcal{M}_3|^2
$ with \begin{align}\Omega_\alpha &\longrightarrow \Omega^{(ab)}_{(n-1,m-1,\ell)},\\
\Omega_\beta &\longrightarrow \Omega^{(c)}_{(n-1,m-1,\ell)}\end{align} both of which are listed in Table~\ref{tab:cases}.~\newline

\paragraph{
Four-mode with selective bilinear exchange.}
Four bosonic modes $a,b,c,d$ with the analogous Hamiltonian and bilinear $c$--$d$ coupling
\begin{equation}
\label{eq: selective coupling}
V_4 = g_{4}\left(c d^\dagger + c^\dagger d\right)
\end{equation}
supporting exchange $|n,m,\ell,k\rangle \leftrightarrow |n,m,\ell{-}1,k{+}1\rangle$.~We retain the three Kerr modes and eliminate mode $d$:
\begin{align}
    H_\mathrm{keep} &= \omega_a a^\dagger a - \frac{K_a}{2}a^{\dagger 2}a^2\\ 
    &\quad+ \omega_b b^\dagger b - \frac{K_b}{2}b^{\dagger 2}b^2 + \omega_c c^\dagger c - \frac{K_c}{2}c^{\dagger 2}c^2 \nonumber \\ H_\mathrm{elim} &=
    \omega_d d^\dagger d  \\
    H_\mathrm{int} &= -\chi_{ab}\, a^\dagger a\, b^\dagger b - \chi_{ac}\, a^\dagger a\, c^\dagger c - \chi_{bc}\, b^\dagger b\, c^\dagger c 
    \\&\quad+ g_{4}\left(c d^\dagger + c^\dagger d\right) \nonumber 
\end{align}
The interaction conserves $n=a^\dagger a$ and $m=b^\dagger b$ individually, since $a$ and $b$ do not appear in $\hat H_{\rm int}$; the only nontrivial invariant is $N\equiv n+m+\ell+k=a^\dagger a+b^\dagger b+c^\dagger c+d^\dagger d$. 

\noindent Within a sector of fixed $(n,m)$, setting $A\equiv n$ and $B\equiv m$, the basis $|r;n,m,N\rangle=|A,B,N-A-B-r,r\rangle$ for $r=0,1,\dots,N-A-B$ renders the Hamiltonian tridiagonal. The jump term 
\begin{equation}
    J_{r+1}^{(n,m)}=g_4\sqrt{(N-A-B-r)(r+1)}
\end{equation}
— unlike the three-mode case, no additional invariant $D$ is needed to fix $A$ and $B$ here, since $n$ and $m$ are already individually conserved.~Denoting
\begin{equation}
  G^{(abc)}_{n,m,\ell,k}(z) \longrightarrow G_4(z) =
  \frac{1}{z - \Omega^{(c)}_{(n,m,\ell-1,k)} - \Sigma_4(z)}
\end{equation}
\begin{equation}
  \Sigma^{(abc)}_{n,m,\ell,k}(z) \longrightarrow \Sigma_4(z) =
  \frac{|\mathcal{M}_4|^2}{z - \Omega^{(d)}_{(n,m,\ell-1,k)}},
  \label{eq:G4Sigma4}
\end{equation}
with dressed poles again following from Eq.~\eqref{eq:Greendressed}. This reduction is simpler than the last: the eliminated mode $d$ contributes a single occupation-conditioned pole shifted by the retained mode occupations through the cross-Kerr couplings with $(b, c)$ only as spectators.~The exchange matrix element is $|\mathcal{M}|^2 \longrightarrow |\mathcal{M}_4|^2 
$, with \begin{align}\Omega_{\alpha}&\longrightarrow\Omega^{(c)}_{(n,m,\ell-1,k)}=\Omega^{(abc)}_{(n,m,\ell-1,k)},\\ \Omega_{\beta}&\longrightarrow \Omega^{(d)}_{(n,m,\ell-1,k)}\end{align} listed in Table~\ref{tab:cases}.\newline

\Novelty{
\paragraph{Sequential eliminations.} One major advantage to the Green's-function construction is that subsystem eliminations are composable.~Once a mode has been projected out, the reduced model retains the same structure required to repeat the procedure: within every sector the Hamiltonian remains tridiagonal, and each elimination replaces the removed subsystem by its continued-fraction self-energy.~Multiple interacting modes may therefore be eliminated sequentially.~Applied to the four-mode parent (see: \textbf{Appendix~\ref{app:sequentialelims}.}) and eliminating the linear mode $d$ and then mode $c$ generates the hierarchy, \begin{equation}G_4 (z) \rightarrow G_4^{(s)} (z) \rightarrow G_3(z)\end{equation} where $G_4^{(s)} (z)$ denotes the intermediate response function of the retained ($a,\, b,\, c$) sector after the elimination of $d$ alone.~Each member of this hierarchy inherits the pole structure of the parent Hamiltonian \footnote{reorganized into successively lower-dimensional, occupation-conditioned response functions,} and thus, reductions extend directly to arbitrary chains of modes.}\newline 

\Contribution{%
Consequently, the two-mode prototype serves as a universal template that extends directly to arbitrary hierarchies of nonlinearly intercoupled subsystems~\footnote{In other words, each elimination preserves the same mathematical structure, so the procedure can be repeated recursively without deriving a new formalism for every additional mode. This is to say that the properties of sequential eliminations establishes that the occupation-conditioned Green's function construction remains closed under successive reduction: each elimination produces another Green's function of identical analytical form, preserving the rational pole structure and continued-fraction self-energy and only needs to be diagonalized at the final stage}.~The recursive nature of the construction becomes particularly useful in the following section, where a sequential elimination establishes the connection between the displaced four-mode and the effective three-wave-mixing spectrum.%
}

\section{\\}\vspace{-0.25cm}\label{sec:coherentdisplacement}
\subsection*{THEORETICAL RESULTS II:\\ Coherent Displacements \& activating \\ Higher Order Mixing Channels}

We can now show precisely in what sense the effective three-wave mixing spectrum arises from a four-mode parent
after linear displacement of one mode.~The distinction is that this equivalence does not come from the purely bilinear exchange Hamiltonian $V_0\propto g_4(cd^\dagger+c^\dagger d)$ alone. Rather,
it follows from a quartic interaction
\begin{equation}
\label{eq:fwm}
V_{4\rm WM}
=
g_4
\left(
abc^\dagger d^\dagger
+
a^\dagger b^\dagger c d
\right),
\end{equation}
whose coherent displacement activates the effective cubic
nonlinearity.~As discussed in \textbf{Appendix~\ref{app:sequentialelims}.}, the (coherently displaced) four-mode parent Hamiltonian and the three-mode mixing Hamiltonian generate identical pole structures, and therefore, identical pseudomode representations, with the elimination order leaving the projected (spectral) response invariant.~This equivalence combines the sequential elimination and spectator shifts of the displaced four-mode parent developed throughout Sec.\textbf{~\ref{sec:coherentdisplacement}.}

{%
Next, consider the \textit{coherent displacement} of retained mode $a$ to arise from $d$ being driven by an external coherent field $\varepsilon_d(t)$, coupled linearly to $d$, so the Heisenberg EoM for its expectation value, denoted $
\xi_d:=\langle d\rangle,
$ reads as
$
\dot \xi_d(t)
=
-i\omega_d\,\xi_d(t)
-i\varepsilon_d(t) - F_d 
$ wherein $F_d$ is related to the memory kernel $K_d(t)$ of Eq.~\eqref{eq: kerneleq} for the $d$-channel — i.e., the same object whose Laplace transform defines
the self-energy $\Sigma_4(z)$ of Eq.~\eqref{eq:G4Sigma4}.~In the stiff-pump regime the drive is stationary, $\varepsilon_d(t)=\varepsilon_d\,
e^{-i\omega_p t}$, and the steady-state coherent amplitude follows
directly from the dressed pseudomode Green's function evaluated at
the pump frequency, $\xi_d
=
-i\,\varepsilon_d\,G_d(\omega_p)$,
using the pole expansion of Eq.~\eqref{eq: decomposition}.~Thus, $\xi_d$ is fixed entirely
by the occupation-conditioned poles and residues of the eliminated
sector's own Green's function rather than a postulated classical displacement.
}

We now displace the linear drive mode according to
$d=\xi_d+\delta d$, wherein $\xi_d\in\mathbb{C}$ is the coherent
pump amplitude and $\delta d$ denotes the residual quantum
fluctuation. Substitution into the parent interaction in \eqref{eq:fwm} gives
$
V_{\mathrm{\tiny
4WM}}
=
g_4
\left[
abc^\dagger(\xi_d^\ast+\delta d^\dagger)
+
a^\dagger b^\dagger c(\xi_d+\delta d)
\right]
$
which factors out to $g_4(\xi_d^\ast abc^\dagger
+
\xi_d a^\dagger b^\dagger c)
+
g_4
(
abc^\dagger\delta d^\dagger+
a^\dagger b^\dagger c\,\delta d
)$.~Therefore, the coherent component activates the effective
three-wave-mixing interaction seen in Eq.~\eqref{eq: mixing scheme coupling} with the effective coupling $g_3\longrightarrow
g_4\xi_d^\ast$ up to the phase convention for $\xi_d$.~The remaining terms vanish under $\delta d\to0$.

~The displacement converts the cross-Kerr couplings involving
$d$ into static Stark shifts with the identification of $\mu\in\{a,b,c\}$ in 
\begin{equation}
g_3=g_4\xi_d^\ast,
\qquad
\tilde\omega_\mu=\omega_\mu-\chi_{\mu}|\xi_d|^2 .
\end{equation}
These shifts are absorbed into occupation-conditioned
transition frequencies of the effective three-mode Hamiltonian. In the stiff-pump limit, the occupation of mode $d$ is sharply peaked
around $k\simeq |\xi_d|^2$, so that
$
\sqrt{k+1}\rightarrow |\xi_d|.
$
Consequently, the exchange element reduces as
\begin{align}
\label{eq: exchangelement_coherent}
g_4\sqrt{nm(\ell+1)(k+1)}
&\longrightarrow
g_4|\xi_d|\sqrt{nm(\ell+1)}
\\
&=
|g_3|\sqrt{nm(\ell+1)}.
\end{align}
This is exactly the three-wave-mixing matrix element obtained from
$abc^\dagger+a^\dagger b^\dagger c$.~The self-energy becomes
\begin{equation}
\Sigma_4^{(s)}(z)
\longrightarrow
\frac{
|g_4|^2|\xi_d|^2nm(\ell+1)
}
{
z-\tilde\Omega^{(bc)}_{(n-1,m-1,\ell)}
}
\rightarrow 
\Sigma_3^{(s)}(z)
\end{equation}
wherein $\tilde\Omega^{(bc)}_{(n-1,m-1,\ell)}$ is the (displaced, composite) transition frequency obtained by substituiting $\omega_\mu\to\tilde\omega_\mu$ for $\mu\in\{a,b,c\}$ after all displacements and absorption of the static Stark shifts.~Therefore, in the stiff-pump limit, where quantum fluctuations of the pumped mode $\delta d$ are neglected, the effective Hamiltonian reduces to the three-mode Hamiltonian analyzed in Sec.~\textbf{\ref{sec:mode-elimination}}, with the effective coupling as $g_3=g_4\xi_d^\ast$ and Stark-shifted mode frequencies.~Consequently, the Green's functions, self-energies, and spectral polar decompositions presented in Sec.\textbf{~\ref{sec:background}.} apply under the substitutions $g_3\to g_4\xi_d^\ast$ and $\omega_\mu \rightarrow \tilde{\omega}_\mu$.

\Contribution{
The equivalence we've established above clarifies why the classical displacement treatment employed throughout cQED successfully reproduces the observed parametric interactions.~In a FWM experiment in circuit QED, the pump mode is driven continuously by an off-resonant coherent microwave tone and is therefore treated as the steady-state response of a damped resonator to an external field.~Consequently, the coherent amplitude is chosen as the expectation value of the driven mode,
$
\xi_d=\langle d\rangle,
$
obtained from the retarded linear susceptibility of the pump mode.~For some \textit{weakly nonlinear} resonator, the intracavity population is proportional to the familiar Lorentzian response,
$
|\xi_d |^{-2}
\propto{(\omega_p-\omega_d)^2+\kappa_d^2/4},
$
where $\omega_p$ is the pump frequency and $\kappa_d$ is the linewidth of the pumped mode. The coherent displacement therefore replaces the operator $d$ by the mean field generated by the resonator's linear response, while the remaining operator $ \delta d=d-\xi_d $ describes only the residual quantum fluctuations about this steady state.~In the stiff-pump regime, these fluctuations become negligible compared to the coherent population.~The conventional classical treatment of the pump is therefore not an independent approximation, but the large-coherent-amplitude limit of the same response-function description underlying the Green's-function reduction developed throughout this work.%
}

\section{Conclusions}\label{sec:conclusion}

\Contribution{Recovering the standard (classical limit) treatment of coherent pump displacement not only confirms the faithfulness of the reduction, it also ensures the scheme is suitable for the optimization of parametric device protocols beyond the applicability of the classical treatment of the drive (and many other common assumptions) presenting a guaranteed significant reduction in the memory (and potentially, computational) cost overhead without compromising on the accuracy of the noise model.} 

We expect the framework to be most useful in regimes which are becoming increasingly common in superconducting hardware designs--~Purcell filters, buffer modes, and lossy readout ports strongly coupled ancillary resonators all introduce structured spectral features whose memory timescales are comparable to the gate times they influence~\cite{Solgun2019,Malekakhlagh2020, MalekakhlaghPetrescuTureci2016, Beaudoin2011}--and the displaced four-mode equivalence of Section~\textbf{\ref{sec:coherentdisplacement}.} suggests an applicability to experiments in which a strongly off-resonantly driven nonlinear coupler mediates an effective parametric interaction between protected modes, including two-photon-dissipative cat qubits~\cite{Mirrahimi2014,Leghtas2015,Putterman2025} and SNAIL-based three-wave mixing~\cite{Frattini2021}.~Some limitations should be flagged: the construction assumes a rational spectral decomposition that is finite, with non-rational densities requiring a controlled approximation step~\cite{Pleasance2020,Menczel2024}; the reduction reproduces the retained subsystem's (reduced) dynamics exactly,~but not the joint system-bath state, so observables on the eliminated modes require purification or input-output extensions~\cite{Lambert2019,Menczel2024};~and the closed-form examples cover small mode counts.

Overall, we have generalized Garraway's pseudomodes construction ~\cite{Garraway1997,Garraway1997b,Imamoglu1994} to retained subsystems with arbitrary nonlinear self/cross-Kerr couplings, formulating the reduction as a Dyson equation whose self-energy is the rational continuation of the eliminated sector's response function.~We've demonstrated the construction in closed form for two-, three-, and four-mode Kerr-coupled systems with bilinear and three-wave-mixing interactions, and established a precise equivalence between a displaced four-mode parent Hamiltonian and the effective three-wave-mixing spectrum observed under a (stiff) coherent pump.~The resulting framework provides a nonperturbative and computationally tractable route to modeling the open quantum system in strongly driven, nonlinear regimes essential to operations in the cQED hardware.
 
\begin{acknowledgments}
This material is based upon work supported by the U.S. Department of Energy, Office of Science, Office of Workforce Development for Teachers and Scientists, Office of Science Graduate Student Research (SCGSR) program. The SCGSR program is administered by the Oak Ridge Institute for Science and Education (ORISE) for the DOE. ORISE is managed by ORAU under contract number DESC0014664.~Part of this research was performed at the Institute for Pure and Applied Mathematics (IPAM), which is supported by the National Science Foundation (Grant No. DMS-1925919).~M. G. B and N. D. thank Cody Fan for helpful discussions.
\end{acknowledgments}

\bibliography{references}

\appendix
\raggedbottom 

\section{\,\\Closure \& exactness of pseudomodes \\ for nonlinearly intercoupled subsystems}
\label{app:nonlinear-green-function-derivation}

In this appendix, we prove the claim of Sec.\textbf{~\ref{sec:general_theory}} that every nonlinearly intercoupled subsystem ensemble of the form of Eq.~\eqref{eq: kerrHamiltonian} admits an exact, finite rational polar decomposition.~We derive how the occupation-conditioned self-energy of Sec.~\textbf{\ref{sec:background}} close at every stage of the elimination, establishing the reductions presented in Sec.~\textbf{\ref{sec:mode-elimination}}.

\subsection{Exacting the conditions for closure}\label{app: closure}
The same conditions for general closure that appear in the original pseudomode construction must also hold for the nonlinearly intercoupled sub-system ensemble.~Thus, this appendix derives the exact conditions in which the nonlinear eliminated subsystem ensemble admits a finite pseudomode representation.~We identify the key mechanism by which the Kerr-type examples in the main text satisfy this condition: because the Kerr nonlinearity is diagonal in the natural number basis — it simply shifts the pre-existing transition frequencies depending on the occupation of the surrounding modes.~These \textit{occupation-conditioned} frequencies keep the pole structure finite and rational even though the underlying system is nonlinear, which keeps the pseudomode construction controlled.\newline

\paragraph{Dyson-type cumulant expansion.}%
We begin from a generic partitioned Hamiltonian of the form of Eq.~\eqref{eq: partitioned} with the interaction $\hat H_\mathrm{int}$ written in channel form as
\begin{equation}
    H_{\mathrm{int}}
    =
    \sum_{\alpha}
    S_{\alpha}^{\dagger}F_{\alpha}
    +
    F_{\alpha}^{\dagger}S_{\alpha}.
\end{equation}
Here, \(S_{\alpha}\) acts only on the retained sector and \(F_{\alpha}\) acts only on the eliminated sector.~The Heisenberg EoM \cite{Heisenberg1925, BornJordan1925} for the retained operator \(S_{\alpha}(t)\) follows from Eq.~\eqref{eq:HeisenbergEOM}.~To the second order in the system--environment coupling, but nonperturbatively in the dynamics of \(H_{\mathrm{elim}}\), the bath influence is determined by the retarded response function, 
\begin{equation}
    \chi_{\alpha\beta}^{(R)}(t-t')
    =
    -i\theta(t-t')
    \langle
    [
    F_{\alpha}(t),
    F_{\beta}^{\dagger}(t')
    ]
    \rangle_{\mathrm{elim}}.
\end{equation}
The response function is defined in the standard sense as the causal (retarded) susceptibility generated by operator commutators \cite[pp.~570--573]{Kubo1957}.~The corresponding memory kernel is therefore 
\begin{equation}
    \mathcal{K}_{\alpha\beta}(t-t')
    =
    \chi_{\alpha\beta}^{(R)}(t-t'),
\end{equation}
Thus, the retained subsystem's dynamics in open-form must close to Eq.~\eqref{eq: totalforce} where \(\xi_{\alpha}(t)\) is the noise operator fixed by the eliminated sector state. Taking the one-sided Laplace transform, denoted 
\begin{equation}
\widetilde{D}(\omega)
\equiv 
\int_0^\infty dt\, e^{i\omega t} D(t),
\end{equation}
recovers the inverse susceptibility matrix,
\begin{align}
\Sigma_{\alpha\beta}(\omega)
&:=
\\
\label{app:retained_eq}
\widetilde{\mathcal{K}}_{\alpha\beta}(t-t')&=
\int_{0}^{\infty} dt\,
e^{i\omega t}
\chi_{\alpha\beta}^{(R)}(t).
\end{align}
The reduction of the convolution to an algebraic product follows directly from the convolution theorem for the one-sided Laplace transform \cite[pp.~40--48]{WidderLaplace}, 
The memory kernel in Eq.~\eqref{app:retained_eq} may then be written as
\begin{equation}
I_{\alpha}(t)
=
\sum_\beta
\int_0^t dt'\,
\mathcal{K}_{\alpha\beta}(t-t')S_\beta(t').
\end{equation}
Introducing the change of variables
$
\tau\longrightarrow t-t',
$ and 
$t'\longrightarrow t-\tau,$
gives an expression of the form
\begin{equation}
I_{\alpha}(t)
=
\sum_\beta
\int_0^t d\tau\,
\mathcal{K}_{\alpha\beta}(\tau)
S_\beta(t-\tau).
\end{equation}
Taking the one-sided Laplace transform then yields
\begin{align}
\widetilde{I_\alpha}(\omega)
&=
\sum_\beta
\int_0^\infty dt\, e^{i\omega t}
\int_0^t d\tau\,
\mathcal{K}_{\alpha\beta}(\tau)
S_\beta(t-\tau)
\\
&=
\nonumber 
\sum_\beta
\int_0^\infty d\tau\,
\mathcal{K}_{\alpha\beta}(\tau)e^{i\omega\tau}
\int_0^\infty du\,
e^{i\omega u}
S_\beta(u),
\end{align}
where in the second line we exchanged the order of integration and defined
$
u \longrightarrow t-\tau.
$ 
The convolution therefore factorizes and admits a time-convolutionless cumulant geometric series expansion (see \cite[pp.~357--364]{Breuer2002}
\begin{equation}
\widetilde{I_\alpha}(\omega)
=
\sum_\beta
\Sigma_{\alpha\beta}(\omega)
S_\beta(\omega),
\end{equation}
with Eq.~\eqref{app:retained_eq}.~The memory kernel equation becomes algebraic in frequency space because temporal convolutions map to multiplication under the Laplace transformation (see also: \cite{Keefe2025}). The convolution thus becomes
\begin{equation}
\label{eq:algebraic}
    \int_{0}^{t}
    dt'\,
    \mathcal{K}_{\alpha\beta}(t-t')S_{\beta}(t')
    \;
    \longrightarrow
    \;
    \Sigma_{\alpha\beta}
    (\omega)S_{\beta}(\omega).
\end{equation}
The retained Green function (and it's Laplace transform by the linearity of the operation) satisfies the Dyson equation \cite{Dyson1949, FetterWalecka}
$
    G^{-1}(\omega)
    =
    G_{0}^{-1}(\omega)
    -
    \Sigma(\omega)
$
as in Eq.~\eqref{eq:Dyson_equation}, wherein \(G_{0}\) is the Green's function generated by \(H_{\mathrm{keep}}\) alone.~The Dyson equation is, in essence, a bookkeeping device: it separates the exact dynamics of a subsystem into a \enquote{bare} piece, $G_0$, that the subsystem would have on its own, and a single correction term, the self-energy $\Sigma(\omega)$, that absorbs everything the eliminated environment does to it. The two combine additively only after an inversion, $G^{-1} = G_0^{-1} - \Sigma$. 
~This is the precise sense in which all eliminated-sector effects enter through the self-energy.~For some nonlinear eliminated sector the retarded response is formally identified as
\begin{equation}
    G_{\alpha\beta}^{(R)}(t)
    =
    -i\theta(t)
    \left\langle
    \left[
    F_{\alpha}(t),
    F_{\beta}^{\dagger}(0)
    \right]
    \right\rangle_{\mathrm{elim}},
\end{equation}
with
\begin{equation}
    F_{\alpha}(t)
    =
    e^{iH_{\mathrm{elim}}t}
    F_{\alpha}
    e^{-iH_{\mathrm{elim}}t}.
\end{equation}
Inserting a complete set of eigenstates
$
    H_{\mathrm{elim}}\left|r\right\rangle
    =
    E_{r}\left|r\right\rangle
$
gives the Lehmann representation in Eq.~\eqref{eq: Lehmann}~\cite{Lehmann1954}.
\begin{widetext}
\begin{equation}
\label{eq: Lehmann}
    G_{\alpha\beta}^{(R)}(\omega)
    =
    \sum_{r,s}
    p_{r}
    \left[
    \frac{
    \left\langle r\right|F_{\alpha}\left|s\right\rangle
    \left\langle s\right|F_{\beta}^{\dagger}\left|r\right\rangle
    }
    {\omega-(E_{s}-E_{r})+i0^{+}}
    -
    \frac{
    \left\langle r\right|F_{\beta}^{\dagger}\left|s\right\rangle
    \left\langle s\right|F_{\alpha}\left|r\right\rangle
    }
    {\omega+(E_{s}-E_{r})+i0^{+}}
    \right],
\end{equation}
\end{widetext}

\paragraph{Remark: On occupation-conditioned transitions.} \!\!\!In the Lehmann representation in Eq.~\eqref{eq: Lehmann},  \(|r\rangle\) and \(|s\rangle\) are many-body eigenstates of the eliminated subsystem with energies \(E_r\) and \(E_s\), \(p_r\) denotes the occupation probability of state \(|r\rangle\), and \(F_\alpha\) are the eliminated-sector operators coupled to the retained subsystem.~This representation makes explicit that the poles of the Green's function correspond to transition frequencies between many-body eigenstates.~Consequently, nonlinear subsystems do possess exact response functions, but these are generally many-body Green's functions containing large or continuous sets of transition poles rather than the finite rational structures characteristic of simple linear reservoirs.~Eq.~\eqref{eq: Lehmann} makes one subtle point explicit: nonlinear subsystem ensembles can have rational Green's functions, but they are generally many-body Green's functions where \(H_{\rm elim}|r\rangle = E_r |r\rangle\) defines the exact many-body eigenstates and eigenenergies of the eliminated sector. The matrix elements \(\langle r|F_\alpha|s\rangle\) therefore quantify the strength of transitions between many-body eigenstates fixed by the coupling.~The infinitesimal term \(i0^+\) imposes the retarded causal boundary condition and fixes the analytic continuation of the response function into the complex frequency plane.~The retarded response (and therefore, its Green's function) is constructed from all energetically allowed transitions of the eliminated subsystem, with poles located at the transition frequencies $ \omega = \pm(E_s-E_r) $.~So far, nothing about this construction of the pseudomode is new; it is the same equation used throughout the literature \cite{Dyson1949, FetterWalecka, Economou2006}, and the self-energy language of this kind already underlies the Fano-diagonalization \cite{Menczel2024} form of the original pseudomode construction \cite{Dalton2001, Pleasance2020}.~What has not been shown is that this equation survives intact when \(H_{\mathrm{keep}}\) is itself nonlinear. Every existing pseudomode derivation of this type implicitly assumes a linear retained subsystem, so that $-i[S_\alpha, H_{\mathrm{keep}}]$ reduces trivially to $-i\Omega_\alpha S_\alpha$ for a fixed frequency $\Omega_\alpha$ and the Laplace transform of the coherent term is immediate.~For the Kerr-type $H_{\mathrm{keep}}$ considered throughout this work, that reduction fails as stated:~the commutator carries an operator-valued coefficient proportional to the retained mode's own occupation, so no single constant $\Omega_\alpha$ exists globally.~The Dyson equation above therefore holds true, but only once this obstruction has been resolved — which requires projecting onto the fixed-occupation sector before the (one-sided) Laplace transform is taken, such that the nonlinear coefficients become a genuine c-number shift within each sector.~We carry out this step explicitly in the following passages, showing precisely how and why closure survives, identifying the occupation-conditioned frequencies that play the role of $\Omega_\alpha$ once the reduction is done sector by sector.\newline

\paragraph{Closing the equation of motion.}%
Eq.~\eqref{eq:algebraic} establishes the memory term in Eq.~\eqref{eq: totalforce} becomes algebraic in frequency space, $\Sigma_{\alpha\beta}
(\omega)S_\beta(\omega)$ for an arbitrary $H_\mathrm{red}$.~This step used only the linearity of $H_\mathrm{int}$ in the eliminated operators $F_\alpha$ and is therefore exact regardless of the structure of $H_\mathrm{red}$.~To close Eq.~\eqref{eq: totalforce} into the Dyson equation for the retained system's Green's function, we must additionally transform the coherent term $-i[S_\alpha(t), H_\mathrm{red}]$, and this step is \emph{not} unconditional: it requires that this commutator be proportional to $S_\alpha$ itself with a $c$-number coefficient.~For a general (possibly nonlinear) $H_\mathrm{red}$ this proportionality fails identically. Consider the Kerr Hamiltonian in Eq.~\eqref{eq: kerrHamiltonian} restricted to a single retained mode $a$,
\begin{equation}
    H_\mathrm{red}
    =
    (\omega_a - \Delta_a)\,\hat a^\dagger \hat a
    -
    \frac{K_a}{2}
    (\hat a^\dagger \hat a)^2.
\end{equation}
Direct evaluation of the commutator gives
\begin{equation}
\label{app: nonlinear_commutator}
    [\hat a, H_\mathrm{red}]
    =
    \Big[
    (\omega_a - \Delta_a)
    -
    K_a\, \hat a^\dagger \hat a
    \Big]\hat a,
\end{equation}
which is proportional to $\hat a$ only up to an operator-valued coefficient, $K_a \hat a^\dagger \hat a$, rather than a $c$-number shift.~Equation~\eqref{app: nonlinear_commutator} therefore does \emph{not} admit a Laplace transform of the simple form $-i\Omega_a S_a(\omega)$ for some constant $\Omega_a$:~the retained sector's own nonlinearity obstructs the (naive) closure of Eq.~\eqref{eq: totalforce} into a linear, frequency-local equation.\newline 

\paragraph{Restoring closure: rojection onto a fixed occupation sector.}%
The obstruction in Eq.~\eqref{app: nonlinear_commutator} is resolved by projecting the retained-sector dynamics onto matrix elements between fixed Fock states, as in the explicit two-mode construction of Section~\textbf{\ref{sec:two_mode}.}~Let $|\Psi(t)\rangle = \sum_{n,m} c_{n,m}(t)\,|n,m\rangle$ denote the state of the enlarged system restricted to the invariant excitation sector, and define $S_\alpha^{(n)}(t) \equiv \langle n|\hat S_\alpha(t)|n\rangle$-type matrix elements accordingly.~In this fixed-occupation subspace, the number operator appearing in Eq.~\eqref{app: nonlinear_commutator} is replaced by its eigenvalue,
$
    \hat a^\dagger \hat a
    \;\longrightarrow\;
    (n+1),
$
so that the commutator becomes proportional to $\hat a$ with a genuine $c$-number coefficient,
\begin{equation}
\label{app: local_freq}
    [\hat a, H_\mathrm{red}]\Big|_{n}
    =
    \Omega_a^{(n)}\,\hat a,
    \quad
    \Omega_a^{(n)}
    \equiv
    (\omega_a - \Delta_a) - K_a\, n.
\end{equation}
This is precisely the (occupation-conditioned) transition frequencies introduced in Eq.~\eqref{eq:energy_two_mode} of the main text. The projection onto a fixed occupation sector is the operation that converts the retained sector's intrinsically nonlinear commutator into a linear, frequency-local one (as in the finite polar decomposition), order by order in occupation.\newline 

\paragraph{Assembling the Dyson equation.}~With Eq.~\eqref{app: local_freq} in hand, the full equation of motion, Eq.~\eqref{eq: totalforce}, restricted to sector $n$ reads
\begin{align}
    \dot S_\alpha^{(n)}(t)
    &=
    -i\,\Omega_\alpha^{(n)}\, S_\alpha^{(n)}(t)
    \\
    \nonumber
    &+
    \sum_\beta \int_0^t dt'\,
    \mathcal K_{\alpha\beta}(t-t')\, S_\beta^{(n)}(t')
    +
    \xi_\alpha(t).
\end{align}
Taking the one-sided Laplace transform from both sides and using the transform of the time derivative,
\begin{equation}
    \widetilde{\dot S_\alpha^{(n)}}(\omega)
    =
    -i\omega\, S_\alpha^{(n)}(\omega)
    -
    S_\alpha^{(n)}(0),
\end{equation}
together with Eq.~\eqref{app:retained_eq} for the memory term, and absorbing the initial-condition contribution $S_\alpha^{(n)}(0)$ into the noise term $\xi_\alpha(\omega)$, gives
\begin{align}
    -i\omega\, S_\alpha^{(n)}(\omega)
    &=
    -i\,\Omega_\alpha^{(n)}\, S_\alpha^{(n)}(\omega)
    \\&+
    \sum_\beta
    \Sigma_{\alpha\beta}^{(n)}(\omega)\,
    S_\beta^{(n)}(\omega)
    +
    \xi_\alpha(\omega).
\end{align}
Collecting terms proportional to $S_\alpha^{(n)}(\omega)$,
\begin{align}
    &\Big[
    \omega - \Omega_\alpha^{(n)}
    -
    \Sigma_{\alpha\alpha}^{(n)}(\omega)
    \Big]
    S_\alpha^{(n)}(\omega)
    \\&\quad\quad\quad=
    \sum_{\beta\neq\alpha}
    \Sigma_{\alpha\beta}^{(n)}(\omega)\, S_\beta^{(n)}(\omega)
    +
    i\,\xi_\alpha(\omega).
\end{align}
Identifying the bracketed prefactor as the inverse retained Green's function, and separating off the bare (uncoupled) piece $\omega - \Omega_\alpha$, 
\begin{equation}
\label{app: dyson_final}
    \big[\widetilde G^{(n)}(\omega)\big]^{-1}
    =
    \big[\widetilde G_0^{(n)}(\omega)\big]^{-1}
    -
    \Sigma^{(n)}(\omega),
\end{equation}
this is exactly the Dyson equation as stated in the main text 
with the identification of $\big[\widetilde G_0^{(n)}(\omega)\big]^{-1} = \omega - \Omega_\alpha^{(n)}$ generated by $H_\mathrm{red}$ alone \emph{at a fixed occupation} of the quantity ($n$), wherein $\Sigma^{(n)}(\omega) \equiv \Sigma_{\alpha\alpha}^{(n)}(\omega)$ carries the complete influence of the eliminated sector through Eq.~\eqref{eq:algebraic}. Eq.~\eqref{app: dyson_final} therefore leads to Eq.~\eqref{eq:Dyson_equation} of the main text, and specializes to Eq.~\eqref{eq: greens_polar} for a single retained channel $\alpha$ and to the explicit two-mode result of Eqs.~\eqref{eq:green_2mode}-\eqref{eq:selfenergy_2mode} upon substituting $\Omega_\alpha^{(n)} \to \Omega^{(a)}_{(n-1,m)}$ and $\Sigma^{(n)}(\omega)\to\Sigma^{(a)}_{(n,m)}(\omega)$.~We emphasize that the sector label ($n$) (or equivalently, the general ($\mu$) in the notation of Sec\textbf{~\ref{sec:general_theory}.}) is not negotiable: Eq.~\eqref{app: nonlinear_commutator} indicates that the retained-sector Green's function given by Eq.~\eqref{app: dyson_final} is well-defined only once the projection in Eq.~\eqref{eq: Lehmann} has been performed, since $\Omega_\alpha^{(n)}$ only becomes a genuine constant within the fixed-occupation sector ($n$).~In the linear limit, $K_a \to 0$, Eq.~\eqref{app: local_freq} becomes occupation-independent, $\Omega_a^{(n)} \to \omega_a - \Delta_a$ for all $n=0,1,2\dots$, the sector label becomes superfluous, and Eq.~\eqref{app: dyson_final} reduces identically to the standard (linear) Dyson-type equation for the retained sector Green's function.\newline


\paragraph{The conditions for closure.} More generally, if the eliminated nonlinear subsystem has discrete dressed eigenstates \(\left|r\right\rangle\), the projected response takes the form
\begin{equation}
    \Sigma_{\alpha\beta}^{(\mu)}(\omega)
    =
    \sum_{r,s}
    \frac{
    \mathcal{M}^{(rs)}_{\alpha \beta}
    }
    {
    \omega-(E_{s}-E_{r})+i\kappa_{rs}/2
    }
    +
    \cdots ,
\end{equation}
in $\mathcal{M}^{(rs)}_{\alpha \beta}:= p_{r}\,
    F_{\alpha}^{(rs)}
    F_{\beta}^{(sr)\ast}$ denotes \(F_{\alpha}^{(sr)}
=
\left\langle r\right|F_{\alpha}\left|s\right\rangle\). If only finitely many transitions have appreciable spectral weight, then truncating to those transitions gives
\begin{equation}
    \Sigma_{\alpha\beta}^{(\mu)}(\omega)
    \simeq
    \sum_{\ell}
    \frac{
    \Lambda_{\alpha\beta}^{(\ell)}
    }
    {\omega-z_{\alpha\beta}^{(\ell)}}.
\end{equation}
which is exactly the pseudomodal reduction.~
Within a fixed occupation sector, the self and cross-Kerr terms shift transition frequencies but do not diverge from sub-harmonic perturbations.~This is why the nonlinear Kerr-dressed examples remain rational: the nonlinearity is diagonal in the chosen number basis and only conditions the pole locations on the spectator occupations. It does not generate an infinite (diverging) continuum inside the fixed sector after diagonalization.

\subsection{Closure \& exactness of the two-mode prototype}\label{app: closuretwomode}
This appendix makes the closure conditions used for the two-mode prototype from Sec.\textbf{~\ref{sec:mode-elimination}.}
explicit. Given $
N=a^\dagger a+b^\dagger b
$,~the number-conserving condition for the retained sub-system, $H_\mathrm{red}$, is $[H_\mathrm{red},N]=0$. Thus, 
\begin{equation}
\mathcal{H}
=
\bigoplus_{N=0}^{\infty}
\mathcal{H}_{N},
\quad
\mathcal{H}_{N}
=
\mathrm{span}
\left\{
|r,N-r\rangle
\right\}_{r=0}^{N}.
\end{equation}
This is the first closure condition: time evolution generated by \(H\) cannot leave a fixed-\(N\) sector.~The diagonal energy of the basis state \(|n,m\rangle\) is Eq.~\eqref{eq: energy_twomode} and the exchange term acts as
\begin{gather}
a^\dagger b|n,m\rangle
=
\sqrt{(n+1)m}
|n+1,m-1\rangle,
\\
b^\dagger a|n,m\rangle
=
\sqrt{n(m+1)}
|n-1,m+1\rangle.
\end{gather}
Thus \(H\) connects only neighboring states inside the same fixed-\(N\) chain:
$
\nonumber
|n,m\rangle
\longleftrightarrow
|n-1,m+1\rangle
\longleftrightarrow
|n-2,m+2\rangle
\longleftrightarrow
\cdots .
$
Equivalently, in the fixed-\(N\) basis \(|r;N\rangle=|r,N-r\rangle\),
\begin{align}
H^{(N)}
&=
\sum_{r=0}^{N}
E_{r,N-r}
|r;N\rangle\langle r;N|
\\&\quad\quad+
\sum_{r=0}^{N-1}
J_{r+1}^{(N)}
\Big(
|r+1;N\rangle\langle r;N|
\\&\quad\quad\quad\quad\quad\quad\quad+
|r;N\rangle\langle r+1;N|
\Big),
\end{align}
with
\begin{equation}
J_{r+1}^{(N)}
=
\tilde{g}\sqrt{(r+1)(N-r)}, 
\end{equation}
leads to the second closure condition: the fixed-sector Hamiltonian must be tridiagonal.~For the local elimination used in the main text, one restricts to the single excitation exchange channel
$
|n,m\rangle
\leftrightarrow
|n-1,m+1\rangle .
$
The channel closes exactly as a two-dimensional subspace only if no additional neighboring amplitudes are dynamically populated. Algebraically, this requires either
$
n=1
\;
\mathrm{or}
\;
m=0,
$
or a controlled projection onto the selected local transition. In the exact fixed-\(N\) problem, additional states are present unless the chain has dimension two.
~Let
$
P
=
|n,m\rangle\langle n,m|,
\;
Q
=
1-P.
$
The exact projected Green function is obtained from the Feshbach identity,
$
G_{P}(z)
=
P(z-H)^{-1}P
=
(
z-E_{n,m}
-
\Sigma_{P}(z)
)^{-1},
$
with exact self-energy
$
\Sigma_{P}(z)
=
PHQ
(z-QHQ)^{-1}
QHP.
$
Because \(H^{(N)}\) is tridiagonal, this self-energy is a continued fraction. If \(P=|r;N\rangle\langle r;N|\), then
\begin{equation}
\Sigma_{r}^{(N)}(z)
=
\Sigma_{r,-}^{(N)}(z)
+
\Sigma_{r,+}^{(N)}(z),
\end{equation}
where
\begin{gather}
\Sigma_{r,+}^{(N)}(z)
=
\frac{
\left(J_{r+1}^{(N)}\right)^2
}{
z-E_{r+1,N-r-1}
-
\Sigma_{r+1,+}^{(N)}(z)
},
\\
\Sigma_{r,-}^{(N)}(z)
=
\frac{
\left(J_{r}^{(N)}\right)^2
}{
z-E_{r-1,N-r+1}
-
\Sigma_{r-1,-}^{(N)}(z)
},
\end{gather}
with 
\begin{equation}
\Sigma_{0,-}^{(N)}(z)=0,
\qquad
\Sigma_{N,+}^{(N)}(z)=0.
\end{equation}
being the boundary conditions.~This is the exact closure of the two-mode nonlinear problem in a fixed excitation sector. The single-pole expression used in the local prototype follows when only one neighboring channel contributes. For the transition in %
$
|n,m\rangle
\rightarrow
|n-1,m+1\rangle,
$
the closure assumptions for the two-mode prototype are simply 
$
[H,N]=0,
$
so that the dynamics remains inside a finite fixed-\(N\) sector; $H^{(N)}
\ \mathrm{is\ tridiagonal}$; so that the exact projected self-energy is a continued fraction; and
\begin{equation}
\Sigma_{r}^{(N)}(z)
\rightarrow
\frac{
\left(J_{r}^{(N)}\right)^2
}{
z-E_{r-1,N-r+1}
}
\end{equation}
is intentionally used as a controlled local projection.~The nonlinear Kerr terms do not destroy closure because they are diagonal in the number basis~\cite{Nigg2012}. They only replace the local oscillator frequencies by occupation-conditioned transition frequencies. 

\section{\,\\Non-Markovian quantum master equation \\ and time-convolutionless expansion}\label{app:redfield}

We will now derive the reduced non-Markovian master equation in Eq.~\eqref{eq:master_prime} from the microscopic treatment of Sec.~\textbf{\ref{sec: microscopic_treatment}.} via the time-convolutionless expansion of the auxiliary pseudomode model.~We show that this derivation recovers the channel-resolved decay kernels $\kappa_j(t)$ of the reductions in Sec.~\textbf{\ref{sec: auxiliary_equation}.} and justifies the form of the master-equation implicitly assumed throughout Sec.\textbf{~\ref{sec:mode-elimination}.}

\subsection{Time-convolutionless expansion of the Redfield-type non-Markovian quantum master equation}


~Let $H^\mu$ denote the retained sub-system.~We represent the dynamics of $H^\mu$ via the master equation in its time convolutionless (TCL) form,
$
\dot \rho_\mu(t)
=
\mathcal{K}(t)\rho_\mu(t),
$
where $\mathcal{K}(t)$ is the TCL generator.~Applying the Born, but not the Markov limit yields the Redfield form, 
\begin{equation}
\dot 
\rho_\mu(t)
=
-i[H^\mu+H_{\mathrm{LS}}^\mu,\rho_\mu(t)]
+
\mathcal{D}[\rho_\mu(t)],
\end{equation}
with
\begin{align}
H_{\mathrm{LS}}
&=
\sum_{\alpha,\beta}
S_{\alpha\beta}(\omega)
A_{\alpha}^{\dagger}(\omega)
A_{\beta}(\omega'),
\end{align}
denoting the Lamb-shifted Hamiltonian as $H_{\mathrm{LS}}^\mu$ and its corresponding TCL dissipator
\begin{align}
\mathcal{D}[\rho_\mu(t)]
&=
\sum_{\alpha,\beta}
\gamma_{\alpha\beta}(\omega)
(
A_{\alpha}(\omega)\rho A_{\beta}^{\dagger}(\omega')
\\&\quad\quad\quad\quad\quad\quad-
A_{\alpha}^{\dagger}(\omega')
A_{\beta}(\omega)\rho
)
+\mathrm{h.c.}
\end{align}
and where $A_{\alpha}(\omega)$ denotes the eigenvalue decomposition
\begin{gather}
A(\omega)
=
\sum_{\epsilon' - \epsilon =\, \omega}
\Pi(\epsilon) A\, \Pi(\epsilon')
\end{gather}
The noise and dissipation kernels appearing in the (non-Markovian) master equation are encoded with the symmetrized~\cite{Kubo1957} reservoir autocorrelation function,
$
f(\tau)\equiv\mathrm{tr}_B\{B(t)B^{\dagger}(t-\tau)\rho_B\}e^{i\omega_0(t-\tau)},
$
\enquote{symmetrized} meaning one-sided, hence $t\geq0$ in Fourier space 
\begin{equation}\Gamma_{\alpha\beta}(\omega)=f_{\alpha\beta}^{(\mu)}(\tau)
=
\int d\omega \,
f_{\alpha\beta}^{(\mu)}(\omega)
e^{i(\omega-\omega_\mu)\tau}\end{equation} 
and its inverse, 
\begin{equation}
\Gamma_{\alpha\beta}^{(\mu)}(\omega)
=
\int_{0}^{\infty} d\tau \,
e^{i\omega\tau}
f_{\alpha\beta}^{(\mu)}(\tau).\end{equation} 
Separating into $\Gamma_{\alpha\beta}(\omega)$ the Cauchy principal value $\mathcal{P}$ of its real and imaginary parts and passing to the continuum limit, maps $\Gamma_{\alpha\beta}(\omega)
\longrightarrow
2\pi J_{\mu}(\omega)$, 
\begin{equation}
    \Gamma_{\alpha\beta}^{(\mu)}(\omega)
=
\frac{\gamma_{\alpha\beta}^{(\mu)}(\omega)}{2}
+
i\,S_{\alpha\beta}^{(\mu)}(\omega).
\end{equation}
and thus yields the symmetric $[\Gamma_{\alpha\beta}(\omega)
+
\Gamma_{\alpha\beta}^{*}(\omega)]/2$ and anti-symmetric $\Gamma_{\alpha\beta}(\omega)
-
\Gamma_{\alpha\beta}^{*}(\omega)/2$ correlators.~The coefficients are written in terms of the spectral density,   
\begin{align}
2\gamma_{\alpha\beta}^{(2)}(\omega)
&=
2\pi
J_{\alpha\beta}^{(2)}(\omega_{\alpha}+\omega),
\\
S_{\alpha\beta}^{(2)}(\omega)
&=
\mathcal{P}
\int d\omega'\,
\frac{
J_{\alpha\beta}^{(2)}(\omega')
}{
\omega_{\alpha}+\omega-\omega'
}.
\end{align}
Here, $\mathcal{P}$ denotes the Cauchy principal value, which
regularizes the singularity at $\omega'=\omega_\alpha+\omega$.
Explicitly, in
\begin{align}
\mathcal{P}
\int d\omega'
\frac{J_{\alpha\beta}^{(2)}(\omega')}
{\omega_\alpha+\omega-\omega'}
\end{align}
$\mathcal{P}$ denotes
\begin{align}
\lim_{\epsilon\to 0^+}
\Big[
\int_{-\infty}^{\omega_\alpha+\omega-\epsilon}
+
\int_{\omega_\alpha+\omega+\epsilon}^{\infty}
\Big]
d\omega'
\frac{J_{\alpha\beta}^{(2)}(\omega')}
{\omega_\alpha+\omega-\omega'}.
\end{align}
This is the quantity that generates the Lamb-shift contribution while excluding the diverging (on-resonance) term.~Then, 
$
2 f_{\alpha\beta}^{(\mu)}(t)
=
\Phi_{\alpha\beta}^{(\mu)}(t)
+
i\Psi_{\alpha\beta}^{(\mu)}(t),
$
where 
\begin{align}
f_{\alpha\beta}^{(\mu)}(t)
&=
\int d\omega'\,
J_{\alpha\beta}^{(\mu)}(\omega')
\exp[
i(\omega_\alpha^{(\mu)}-\omega')t
].
\end{align}
The kernel
$f_{\alpha\beta}^{(\mu)}(t)$ may be expressed through the
pseudomode pole expansion introduced in ~\textbf{\ref{sec: pseudomode_construction}.}~Writing the effective spectral density of the eliminated
sector as 
\begin{align}
J_{\alpha\beta}^{(\mu)}(\omega)
=
\sum_{\ell}
\frac{r_{\alpha\beta,\ell}^{(\mu)}}
{\omega-z_{\ell}^{(\mu)}},
\quad
z_{\ell}^{(\mu)}
=
\Omega_{\ell}^{(\mu)}
-\frac{i}{2}\kappa_{\ell}^{(\mu)},
\end{align}
the corresponding memory kernel becomes
\begin{align}
f_{\alpha\beta}^{(\mu)}(t)
=
-i
\sum_{\ell}
r_{\alpha\beta,\ell}^{(\mu)}
e^{-i\Omega_{\ell}^{(\mu)} t}
e^{-\kappa_{\ell}^{(\mu)} t/2}.
\end{align}
hence
$
\Phi_{\alpha\beta}^{(\mu)}(t)
=
2\,\mathrm{Re}
[
f_{\alpha\beta}^{(\mu)}(t)
],
\; 
\Psi_{\alpha\beta}^{(\mu)}(t)
=
2\,\mathrm{Im}
[
f_{\alpha\beta}^{(\mu)}(t)
],
$
so that the TCL coefficients below inherit their frequency dependence directly from the pseudomode
frequencies $\Omega_{\ell}^{(\mu)}$ and linewidths $\kappa_{\ell}^{(\mu)}$ as defined in Sec.~\textbf{\ref{sec: pseudomode_construction}.}~To the 2nd order coupling, they are  
\begin{align}
\gamma_{\alpha\beta}^{(2,\mu)}(t)
&=
\int_0^t dt_1\,
\Phi_{\alpha\beta}^{(\mu)}(t-t_1),
\\
S_{\alpha\beta}^{(2,\mu)}(t)
&=
\int_0^t dt_1\,
\Psi_{\alpha\beta}^{(\mu)}(t-t_1).
\end{align} 
and to the 4th order in Eq.'s~\eqref{eq: 4thorder1}-\eqref{eq: 4thorder2} where $\tau\equiv t_1$, recovering the Born-Markov limit as $t\longrightarrow\infty$, 
\begin{align}
\gamma_{\alpha\beta,M}^{(\mu)}
&=
\int_0^\infty ds\,
\Phi_{\alpha\beta}^{(\mu)}(s),
\\
S_{\alpha\beta,M}^{(\mu)}
&=
\int_0^\infty ds\,
\Psi_{\alpha\beta}^{(\mu)}(s).
\end{align}
In contrast to the enlarged Markovian model, it modifies the higher order channels in the reduced description.

\begin{widetext}
\begin{align}
\label{eq: 4thorder1}
\gamma_{\alpha\beta}^{(4,\mu)}(t)
=
\gamma_{\alpha\beta}^{(2,\mu)}(t)+
\frac{1}{2}
\sum_{\eta}
\int_0^t dt_1
\int_0^{t_1} dt_2
\int_0^{t_2} dt_3
&
[\;
\Phi_{\alpha\eta}^{(\mu)}(t-t_2)
\Phi_{\eta\beta}^{(\mu)}(t_1-t_3)
+
\Phi_{\alpha\eta}^{(\mu)}(t-t_3)
\Phi_{\eta\beta}^{(\mu)}(t_1-t_2)
\nonumber
\\
&
-
\Psi_{\alpha\eta}^{(\mu)}(t-t_2)
\Psi_{\eta\beta}^{(\mu)}(t_1-t_3)
-
\Psi_{\alpha\eta}^{(\mu)}(t-t_3)
\Psi_{\eta\beta}^{(\mu)}(t_1-t_2)
\;],
\\ \nonumber\\ \label{eq: 4thorder2}
S_{\alpha\beta}^{(4,\mu)}(t)
=
S_{\alpha\beta}^{(2,\mu)}(t)+
\frac{1}{2}
\sum_{\eta}
\int_0^t dt_1
\int_0^{t_1} dt_2
\int_0^{t_2} dt_3&
[\;
\Psi_{\alpha\eta}^{(\mu)}(t-t_2)
\Phi_{\eta\beta}^{(\mu)}(t_1-t_3)
+
\Phi_{\alpha\eta}^{(\mu)}(t-t_2)
\Psi_{\eta\beta}^{(\mu)}(t_1-t_3)
\\ \nonumber
&
+
\Psi_{\alpha\eta}^{(\mu)}(t-t_3)
\Phi_{\eta\beta}^{(\mu)}(t_1-t_2)
+
\Phi_{\alpha\eta}^{(\mu)}(t-t_3)
\Psi_{\eta\beta}^{(\mu)}(t_1-t_2)
\;].
\end{align}
\end{widetext}




For any retained mode ($\mu$) obtained from any reduction in Table~\ref{tab:cases}, the projected self-energy $\Sigma^{(\mu)}(\omega)$ closes onto exactly two dissipative channels: the native single-exchange channel of the reduction ($L_1 = \hat a_\mu$), and---once the eliminated sector contains a displacement per pump mode as in Sec.\textbf{~\ref{sec:coherentdisplacement}.}---a higher-order channel activated by that displacement ($L_2 = \hat a_\mu^2$).~For $\mu=\{r\}$, the coupling rates may then be written as dressed composite transition rates. The retained transitions consist of the available channels~%
$
(n,m,\ell)\rightarrow(n,m-1,\ell),
$~%
coupled through 
$
|\mathcal{M}_{n,m,\ell}^{(1r)}|
$
while the dressed poles $z_{\alpha\beta}^{(r)}$ are obtained from the composite transition rates for the retained sector, 
$
\Omega_{n,m,\ell}^{(1r)}
$
%
~In addition, the $\hat{H}_\mathrm{FWM}$ interaction activates the composite exchange channel $(n,m,\ell)
\rightarrow
(n,m-1,\ell+2)$, leading to the form of the coupling 
$
|\mathcal{M}_{n,m,l}^{(2s)}|. 
$
\label{eq: weight2}
The composite transition frequency
$
\Omega_{\alpha\beta}^{(2)}\equiv\Omega_{\alpha\beta}^{(n-1,m+2,\ell)}
$
represents the activated transition rates associated with annihilating one photon in the mode $r$, while simultaneously creating two photons in the mode $s$, up to the (spectator) self-Kerr shifts generated by the remaining storage occupation $\ell$, denoted 
$
\Omega_{n,m,\ell}^{(2s)}.
$

Identifying one (pseudo-)mode per channel, 
$
z^{(s1)}_{n,m,\ell}
=
\Omega^{(s1)}_{n,m,\ell}
-
i \kappa_{1}/2
$~and~$
z^{(s2)}_{n,m,\ell}
=
\Omega^{(s2)}_{n,m,\ell}
-
i\kappa_{2}/2,
$
~leads to \begin{equation}
J^{(\mu)}(\omega)
\longrightarrow
J_s^{(s1)}(\omega)
+
J_s^{(s2)}(\omega),
\end{equation} where
\begin{gather}
J_s^{(s1)}(\omega)
=
\sum_{n=0}^{\infty}
\sum_{m=1}^{\infty}
\sum_{\ell=0}^{\infty}
\frac{|\mathcal{M}_{n,m,\ell}^{(1r)}|
}
{
[\omega-\Omega^{(s1)}_{n,m,\ell}]^2
+
[\kappa_{1}
/2]^2
},
\\
J_s^{(s2)}(\omega)
=
\sum_{n=0}^{\infty}
\sum_{m=1}^{\infty}
\sum_{\ell=0}^{\infty}
\frac{|\mathcal{M}_{n,m,l}^{(2s)}|
}
{
[\omega-\Omega^{(s2)}_{n,m,\ell}]^2
+
[\kappa_{2}
/2]^2
},
\end{gather}
the retained-sector only reduced model is Eq.~\eqref{eq:master_prime} with the single-photon channel,
\begin{equation}
\kappa_1(t)
=
2
\sum_{n,m,\ell}
|\Gamma_1|^2
\,
{\rm Re}
\Big[
\frac{
1-
e^{-[\kappa^{(s1)}_{\rm eff}/2+i\Omega^{(s1)}_{n,m,\ell}]t}
}
{
\kappa_{\rm eff}^{(s1)}/2
+
i\Omega^{(s1)}_{n,m,\ell}
}
\Big],
\label{eq: kappa1_eff}
\end{equation}
and
\begin{align}
\kappa_2(t)
&=
2
\sum_{n,m,\ell}
|\Gamma_2|^2
{\rm Re}
\Big[
\frac{
1-
e^{-[\kappa^{(s2)}_{\rm eff}/2+i\Omega^{(s2)}_{n,m,\ell}]t}
}
{
\kappa^{(s2)}_{\rm eff}/2
+
i\Omega^{(s2)}_{n,m,\ell}
}
\Big].
\label{eq: kappa2_eff}
\end{align}
In the long-time Markovian limit, these kernels approach constant rates,
%
%
%
and the (composite) frequencies, $\Omega^{(sj)}_{n,m,\ell}$, linewidths and couplings entering Eqs.~\eqref{eq: poles_solution} are obtained by diagonalizing the (truncated) generator of the eliminated sector; the pole positions of Eq.~\eqref{eq:spectraldensitypseudo} provide the most direct route to the effective decay rates entering the reduced description in Eq.~\eqref{eq:master_prime}.

Following the TCL construction above, the Lamb shift contribution becomes
\begin{equation}
\delta\omega^{\rm LS}_{(\mu)}(n) \equiv S^{(\mu)}_{\alpha}\,\Omega_\alpha^{(\mu)}
= \sum_{\beta} \mathcal{P}\!\!\int d\omega' \,
\frac{J^{(\mu)}_{\alpha\beta}(\omega')}{\Omega_{\alpha\beta}^{(\mu)}-\omega'},
\label{eq:D2}
\end{equation}
the occupation-conditioned bare frequency of Eq.~\eqref{eq: kerrHamiltonian} is renormalized and the Lamb-shifted reduced Hamiltonian is obtained from $\hat H_{\rm red}$ (Eq.~\eqref{eq: reduced_Hamiltonian_definition}) as $\Omega_\alpha^{(n)}\to\widetilde\Omega_\alpha^{(n)} \equiv \Omega_\alpha^{(n)}+\delta\omega_{(\mu)}^{\rm LS}(n)$ by the mode-by-mode substitution, 
\begin{align}
&\hat H^{\rm LS}_{\rm red} =
\sum_{\mu}\Big[\big(\omega_\mu-\Delta_\mu+\delta\omega^{\rm LS}(\hat n_\mu)\big)\hat a_\mu^\dagger\hat a_\mu
\\&\quad\quad\quad\quad-\frac{K_\mu}{2}(\hat a_\mu^\dagger\hat a_\mu)^2\Big]
-\sum_{\mu\neq\nu}\chi_{\mu\nu}\,\hat n_\mu\hat n_\nu.
\label{eq:D3}
\end{align}
Because $\delta\omega^{\rm LS}(n)$ depends on ($n$) only through the same occupation label that already enters $\Omega_\alpha$, Eq.~\eqref{eq:D3} remains diagonal in the Fock basis and preserves the tridiagonal sector structure of Sec~\textbf{~\ref{sec:two_mode}.}---the Lamb shift does not reintroduce off-diagonal coupling and therefore does not spoil the closure conditions defined in \textbf{Appendix~\ref{app:nonlinear-green-function-derivation}.}~For any row of Table~\ref{tab:cases}, the single-exchange decay kernel and its Markov limit take the same form as in the two-mode prototype example, with the matrix element and transition frequency read directly from Table~\ref{tab:cases} where $\Delta\Omega\equiv\Omega_\alpha-\Omega_\beta$ is the row's local detuning and $-2\,{\rm Im}\,\{z_{\alpha\beta}\}$ is the linewidth of the corresponding dressed pole. The row-specific substitutions are:

\begin{table}[h]
\centering
\begin{tabular}{lll}
\hline\hline
Row & $|M_{(S_j)}|^2$ & $\Delta\Omega$ \\
\hline
1 & $S_j=1 \; \; g_{ab}^2\,n(m+1) \quad$        & $\Omega^{(a)}_{(n-1,m)}-\Omega^{(b)}_{(n-1,m)}$ \\
2 & $S_j=1 \; \; g_3^2\,nm(\ell+1) \quad $       & $\Omega^{(ab)}_{(n-1,m-1,\ell)}-\Omega^{(c)}_{(n-1,m-1,\ell)}$ \\
3 & $S_j=1 \; \; g_4^2\,\ell(k+1) \quad$        & $\Omega^{(c)}_{(n,m,\ell-1,k)}-\Omega^{(d)}_{(n,m,\ell-1,k)}$ \\
\hline\hline
\end{tabular}
\caption{Row-specific substitutions for the two-mode prototype couplings, with $|\mathcal{M}|,\Omega_\alpha,\Omega_\beta$ taken from Table~\ref{tab:cases}.}
\label{tab:D1}
\end{table}
\noindent with the sum running over the surviving spectator occupation numbers of the fixed sector (i.e., over the accessible range of $r$ in Eqs.~\eqref{eq: kappa1_eff}-~\eqref{eq: kappa2_eff} at the coupling index $S_j$ corresponding to the interaction).

\section{\,\\Sequential eliminations \& Closure \\ to the Classical Limit}\label{app:sequentialelims}

In this appendix, we give the explicit derivation of the sequential elimination procedure invoked in Sec~\textbf{\ref{sec: pseudomode_construction}.}, applied concretely to the four-mode parent Hamiltonian of Sec~\textbf{\ref{sec:coherentdisplacement}.} Our goal is to show, step by step, that eliminating the linear mode $d$ and then the Kerr mode $c$ produces a well-defined hierarchy of exact Green's functions,
\begin{equation}
G_4(z)
\;\longrightarrow\;
G_4^{(s)}(z)
\;\longrightarrow\;
G_3(z),
\end{equation}
in which each stage retains the same rational, occupation-conditioned pole structure established in Sec\textbf{~\ref{sec:background}.} We first eliminate mode $d$ alone, obtaining the intermediate self-energy and the corresponding Green's function for the retained $(a,b,c)$ sector; we then eliminate mode $c$ from this reduced problem, recovering the two-mode Green's function $G_3(z)$ of Sec~\textbf{~\ref{sec:two_mode}.}~This explicit construction substantiates the composability claim of Sec~\textbf{\ref{sec: pseudomode_construction}.}: that once a subsystem has been eliminated, the resulting reduced problem can be treated by the same continued-fraction machinery as the original, so that no new formalism is required at each successive stage. The derivation therefore confirms that the elimination hierarchy of Fig.~\ref{fig:hiearchy} is not merely schematic, but follows directly and exactly from repeated application of the single-mode elimination template of Sec\textbf{~\ref{sec:two_mode}.}

\subsection{Problem Setup}

The reductions developed in Sec~\textbf{\ref{sec:mode-elimination}} are composable because each fixed-sector Hamiltonian remains tridiagonal before projection, while each elimination replaces the removed subsystem by its exact continued-fraction self-energy.~Consequently, the pseudomodal elimination procedure may be sequentially applied without changing the algebraic structure of the retained Green's function. 
\newline

\paragraph{Sequential elimination of the three-mode mixing Hamiltonian}%
For the three-mode mixing Hamiltonian, conservation of the total excitation number $Q=n+m+2\ell$ and excitation imbalance $D=n-m$ decomposes the Hilbert space into independent $(Q,D)$ sectors spanned by $|r;Q,D\rangle= |A-r,B-r,r\rangle$,
where
\begin{equation}
A=\frac{Q+D}{2},
\qquad
B=\frac{Q-D}{2}.
\end{equation}
\noindent Within each sector the Hamiltonian is tridiagonal,
\begin{align}
&H^{(Q,D)}
=
\sum_r
E_r^{(Q,D)}
|r\rangle\langle r|
\\&\nonumber +
\sum_r
J_{r+1}^{(Q,D)}
(
|r+1\rangle\langle r|
+\mathrm{h.c.}),
\end{align}
with $E_r^{(Q,D)}
=
E_{A-r,B-r,r}$ and 
\begin{align}
J_{r+1}^{(Q,D)}
=
g_3
\sqrt{(A-r)(B-r)(r+1)},
\end{align}
matching the substitutions of Table~I: at $r=\ell$, the sector basis state $|A-r,B-r,r\rangle$ recovers the physical Fock state $|n,m,\ell\rangle$.~Applying the continued-fraction elimination developed to this tridiagonal chain yields, for each sector, the general local Green's function
\begin{equation}
G_r^{(Q,D)}(z)
=
\frac{
1
}{
z-
E_r^{(Q,D)}
-
\Sigma_{r,-}^{(Q,D)}(z)
-
\Sigma_{r,+}^{(Q,D)}(z)
},
\end{equation}
with
\begin{align}
\Sigma_{r,\pm}^{(Q,D)}(z)
&=
\frac{
(J_{r\pm1}^{(Q,D)})^2
}{
z-
E_{r\pm 1}^{(Q,D)}
-
\Sigma_{r\pm1,\pm}^{(Q,D)}(z)
}
\end{align}
subject to the boundary conditions $\Sigma_{0,-}^{(Q,D)}(z)=0$ and $\Sigma_{\min(A,B),+}^{(Q,D)}(z)=0$. Restricting to the physical sector value $r=\ell$ recovers the composite $(a,b)$ retained Green's function $G_3(z)$ and self-energy $\Sigma_3(z)$ of Eqs.~\eqref{eq: G3_greens}-\eqref{eq:G3Sigma3}, since $\Sigma_{r,-}^{(Q,D)}(z)\big|_{r=\ell}$ is precisely the single-pole contribution from the eliminated mode $c$.\newline 

\paragraph{Sequential elimination of the four-mode parent Hamiltonian}%
The four-mode Hamiltonian admits an analogous decomposition.~The modes ($a$) and ($b$) do not appear in $H_{\rm int}$ and are therefore individually conserved: $n=a^\dagger a$ and $m=b^\dagger b$ are good quantum numbers, entering only as Kerr-shifted spectators.~The only nontrivial invariant generated by the bilinear $c$--$d$ coupling is $N\equiv n+m+\ell+k$. Within a sector of fixed $(n,m)$, setting $A\equiv n$ and $B\equiv m$, the basis
$
|r;n,m,N\rangle = |A,B,N-A-B-r,r\rangle
$
for $r=0,1,\dots,N-A-B$ renders the Hamiltonian tridiagonal,
\begin{align}
&H^{(n,m)}
=
\sum_r E_r^{(n,m)} |r;n,m,N\rangle\langle r;n,m,N|
\\&+
\sum_r J_{r+1}^{(n,m)}
\left(
|r{+}1;n,m,N\rangle\langle r;n,m,N|+\mathrm{h.c.}
\right),
\end{align}
with entries $E_r^{(n,m)}
=
E_{A,B,N-A-B-r,r}$ and hopping amplitude
\begin{align}
J_{r+1}^{(n,m)}
=
g_4
\sqrt{(N-A-B-r)(r+1)},
\end{align}
matching Eq.~(91): at $r=k$, the sector basis state recovers the physical Fock state $|n,m,\ell,k\rangle$, since $N-A-B-r\big|_{r=k}=\ell$.~The continued-fraction self-energy therefore applies without modification,
\begin{equation}
G_r^{(n,m)}(z)
=
\frac{
1
}{
z-
E_r^{(n,m)}
-
\Sigma_{r,-}^{(n,m)}(z)
-
\Sigma_{r,+}^{(n,m)}(z)
},
\end{equation}
with
\begin{equation}
\Sigma_{r,\pm}^{(n,m)}(z)
=
\frac{
(J_{r\pm1}^{(n,m)})^2
}{
z-
E_{r\pm1}^{(n,m)}
-
\Sigma_{r\pm1,\pm}^{(n,m)}(z)
},
\end{equation}
subject to $\Sigma_{0,-}^{(n,m)}(z)=0$ and $\Sigma_{N-A-B,+}^{(n,m)}(z)=0$ as the boundary conditions.~Since the mode $d$ occupies the terminal site $r=k$ of this chain and couples only to $c$, eliminating $d$ alone leaves a single dressed pole acting on the retained $(a,b,c)$ sector: restricting to $r=k$ recovers exactly the intermediate response $G_4^{(s)}(z)$ and self-energy $\Sigma_4(z)$ of Eqs.~\eqref{eq: G3_greens}-\eqref{eq:G3Sigma3}.~The subsequent elimination of mode $c$ from this reduced problem is then required to reach the fully retained response, completing the hierarchy $G_4(z)\longrightarrow G_4^{(s)}(z)\longrightarrow G_3(z)$ claimed in Sec~\textbf{\ref{sec:multimodes}.}\textit{c}.\newline 

\paragraph{Proof Sketch.} After proving closure of the quartic interaction, the derivation proceeds in five steps.~First, we construct the local four-mode Green's function directly from the undisplaced parent Hamiltonian.~Second, we apply the displacement transformation to the Hamiltonian itself.~Third, we obtain the effective three-mode Hamiltonian in the stiff-pump limit.~Fourth, we derive the self-energy of that effective Hamiltonian independently.~Finally, we compare the two reduced expressions and identify the conditions under which their Green's functions and pole structures coincide.

\subsection{Closure of the four-wave mixing coupling}

The sequential elimination argument that was made in Sec~\textbf{\ref{sec:multimodes}.}\textit{c} and presented throughout Sec~\textbf{\ref{sec:coherentdisplacement}.} requires
that the displaced four-mode parent Hamiltonian $\hat H_4 = \hat H_{\rm int} + \hat V_{\rm 4WM}$,
defined in Eq.~\eqref{eq:fwm} %
admit an exact finite-rational self-energy as in \textbf{Appendix~\ref{app: closuretwomode}.}~prior to any displacement.~Since $\hat V_{\rm 4WM}$ is a genuinely distinct
coupling from the bilinear $c$--$d$ exchange in Sec~\textbf{\ref{sec:multimodes}.}\textit{b}, closure must be established independently.
\newline 

\paragraph{Conserved quantities.}
Under $\hat V_{\rm 4WM}$, a single application shifts the Fock labels by
$(\Delta n, \Delta m, \Delta \ell, \Delta k) = \mp(1,1,-1,-1)$. $\hat H_{4,0}$
is diagonal in the number basis and contributes no off-diagonal terms,
so the conserved quantities of $\hat H_4$ are exactly those of
$\hat V_{4WM}$. A linear combination $c_n n + c_m m + c_\ell \ell + c_k k$
is conserved iff $c_\ell + c_k = c_n + c_m$, a single constraint on four
coefficients, leaving a three-dimensional space of invariants. We take
as an independent basis
\begin{equation}
D_{ab} \equiv n-m, \quad D_{\ell k} \equiv \ell-k, \quad
N_4 \equiv n+m+\ell+k,
\label{eq:Cc2}
\end{equation}
each of which is readily checked to be invariant.~Note in particular that, in contrast to the \emph{bilinear} four-mode parent of Sec~\textbf{\ref{sec:multimodes}.}\textit{b}--- where $n$ and $m$ are individually conserved because $a,b$ do not appear in $\hat H_{\mathrm{int}}$ --- under $\hat V_{4WM}$ only the combination $D_{ab}=n-m$ survives; $n$ and $m$ separately are not conserved.\newline 

\paragraph{Fixed Sector basis.} The two four-mode Hamiltonians therefore populate different sector structures and should not be treated as sharing a single reduction hierarchy without this distinction made explicit.~Fixing $(N_4, D_{ab}, D_{\ell k})$ and defining
\begin{equation}
A \equiv \frac{N_4 + D_{ab} + D_{\ell k}}{2}, \quad
B \equiv \frac{N_4 - D_{ab} + D_{\ell k}}{2}.
\label{eq:Cc3}
\end{equation}
Solving the three conservation laws for $(n,m,k)$ in terms of a single
free integer $r \equiv \ell$ gives
\begin{equation}
n = A-r, \quad m = B-r, \quad \ell = r, \quad k = r - D_{\ell k},
\label{eq:Cc4}
\end{equation}
so the basis  %
$|r; N_4, D_{ab}, D_{\ell k}\rangle \equiv |A-r,\,B-r,\,r,\,r-D_{\ell k}\rangle$
is in correspondence with the physical Fock states of the sector over
$
r \in \big[\max(0,D_{\ell k}),\ \min(A,B)\big].
\label{eq:Cc5}
$
\newline 

\paragraph{Tridiagonal matrix structure.}
Within the fixed $(N_4,D_{ab},D_{\ell k})$ sector, $\hat V_{\rm 4WM}$ connects only $|r\rangle \leftrightarrow |r+1\rangle$: each application of $\hat V_{\rm 4WM}$ executes exactly the shifted  $(\Delta n,\Delta m,\Delta\ell,\Delta k)=(-1,-1,+1,+1)$ or its conjugate, and $\hat H_{\rm int}$ contributes no off-diagonal elements at all.~No operator in $\hat H_4$ connects $|r\rangle$ to $|r\pm2\rangle$ or beyond in a single step. The sector Hamiltonian is therefore exactly tridiagonal,
\begin{align}
&\hat H^{(N_4,D_{ab},D_{\ell k})} = \sum_r E_r\,|r\rangle\langle r|
\nonumber\\&\quad\quad+ \sum_r J_{r+1}\big(|r{+}1\rangle\langle r| + \mathrm{h.c.}\big),
\label{eq:Cc6}
\end{align}
with jump amplitude
\begin{align}
J_{r+1} &= g_4\sqrt{(A-r)(B-r)(r+1)(r+1-D_{\ell k})} \\
&= g_4\sqrt{n\,m\,(\ell+1)(k+1)}\Big|_{\ell=r},
\label{eq:Cc7}
\end{align}
which reproduces the matrix element $|\mathcal{M}_4|^2$ taken from Eq.~\eqref{eq: jumpterm3} of the
main derivation in Eq.~\eqref{eq: exchangelement_coherent}.\newline 

\paragraph{Exact self-energy and boundary conditions.}
This is now formally identical to the two-mode closure of
\textbf{Appendix~\ref{app: closuretwomode}.}, re-indexed by $r$. The Feshbach-projected resolvent at
site $r$ is
\begin{gather}
G_r(z) = \frac{1}{z - E_r - \Sigma_{r,-}(z) - \Sigma_{r,+}(z)}, \\
\Sigma_{r,\pm}(z) = \frac{J_{r\pm1}^2}{z - E_{r\pm1} - \Sigma_{r\pm1,\pm}(z)},
\label{eq:Cc8}
\end{gather}
subjected to $\Sigma_{r_{\min},-}(z) = 0$ and $\Sigma_{r_{\max},+}(z) = 0$~and~$r_{\min}=\max(0,D_{\ell k}),\, r_{\max}=\min(A,B)$.~Equations~\eqref{eq:Cc6}--\eqref{eq:Cc8} constitute the exact closure
of $\hat H_4$ in a fixed sector, completing the result. In particular, this confirms that $\hat H_4$ admits a finite rational polar decomposition as presented in \textbf{Appendix \ref{app: closure}-\ref{app: closuretwomode}}, with the poles $\{z_r\}$ obtained by diagonalizing the truncated finite tridiagonal matrix~\eqref{eq:Cc6}.\newline 

\paragraph{Relation to the single-pole form.}The self-energy quoted in Eq.~(C18),
\begin{equation}
\Sigma_4^{\rm (s)}(z) = \frac{|g_4|^2\, n m (\ell+1)(k+1)}
{z-\Omega^{(bcd)}_{n-1,m-1,\ell,k}},
\label{eq:Cc10}
\end{equation}
corresponds to keeping only the leading term of $\Sigma_{r,+}(z)$ in
the continued fraction~\eqref{eq:Cc8} and dropping $\Sigma_{r,-}(z)$
entirely. By the boundary condition, $\Sigma_{r,-}(z)$
vanishes exactly, and Eq.~\eqref{eq:Cc10} is exact, whenever the
transition of interest originates at $r=r_{\min}$ --- physically,
whenever mode $c$ carries no excitation prior to the exchange beyond
what is already fixed by the sector label ($\ell = D_{\ell k}$ when
$D_{\ell k}\ge0$, or $\ell=0$ when $D_{\ell k}<0$). For $r>r_{\min}$,
$\Sigma_{r,-}(z)$ is generically nonzero, and Eq.~\eqref{eq:Cc10}
should be understood as the leading term of the finite continued
fraction~\eqref{eq:Cc8} rather than the exact resolvent; the chain
length $r_{\max}-r_{\min}\sim n+m+2\ell$ is set by the (typically
weakly populated) retained-mode occupations, so the omitted terms are
controlled and can be bounded directly from
Eq.~\eqref{eq:Cc8} rather than assumed negligible. When the same two modes eliminated in Sec~\textbf{\ref{sec:multimodes}.}\textit{b}-\textit{c} are eliminated together with the linear drive mode $d$, the retained subsystem is now only the mode $a$. The retained transition frequency of mode $a$ is
\begin{align}
\Omega_\alpha \longrightarrow \Omega^{(a)}_{(n-1,m,\ell,k)}
&=
\omega_a
-K_a(n{-}1)
\\&\quad-\chi_{ab}(m)
-\chi_{ac}(\ell)
. \nonumber
\end{align}
while 
\begin{align}
\Omega_\beta \longrightarrow \Omega^{(bcd)}_{(n-1,m-1,\ell,k)}
&=
\omega_c+\omega_d-\omega_b
\nonumber\\
&\quad
+K_b(m{-}1)-K_c(\ell)
\nonumber\\
&\quad
+\left(\chi_{ab}-\chi_{ac}\right)(n{-}1)
\nonumber\\
&\quad
-\chi_{bc}\left(m-1-\ell\right)
\end{align}
is the composite transition frequency of the eliminated sector.~Applying the same local elimination procedure presented in Sec~\textbf{\ref{sec:multimodes}.}\textit{a} gives
\begin{gather}\nonumber 
G_{3}^{\rm eff}(z) \longrightarrow G_{4}^{(\rm s)}(z)
=
\frac{1}
{
z-\Omega^{(a)}_{(n-1,m,\ell,k)}
-\Sigma_4^{(\rm s)}(z)
},
\\
\label{eq:selfenergy_4mode_two_modes_drive_elim}
\Sigma_{3}^{\rm eff}(z) \longrightarrow \Sigma_{4}^{(\rm s)}(z)
=
\frac{
|g_4|^2 n m(\ell+1)(k+1)
}
{
z-\Omega^{(b cd)}_{(n-1,m-1,\ell,k)}
}.
\end{gather}
Therefore, eliminating two nonlinear modes ($cd$) together with the linear drive mode ($d$) produces a singular composite occupation-conditioned pole.~Here, $(b cd)$ indicates that the eliminated process removes one quantum from ($b$) while creating one quantum in ($c$) and one quantum in the drive mode ($d$).~We utilize this single-pole form in Sec~\textbf{\ref{sec:coherentdisplacement}.} accordingly, restricting to the low-occupation sector of the retained modes with a Kerr nonlinearity in which the occupation-conditioned transition frequencies are exact or near-exact.\newline 

\paragraph{Main takeaways.} Because the Green's functions coincide identically in \(z\), their poles, residues, spectral densities, memory kernels, and pseudomode representations all coincide.~Away from the \enquote{stiff-pump} limit, the correction is generated by vacuum fluctuations and the fluctuation-dependent terms vanish as \(\sqrt{\langle\delta d^\dagger\delta d\rangle}/|\xi_d|\to0\), consistent with \(\sqrt{k+1}\simeq|\xi_d|\) at a large occupation. The logical sequence 
\begin{equation}
H_4 \to H_4^{(\rm s)} \to D^\dagger(\xi_d)H_4D(\xi_d) \to H_{3}^{\mathrm{eff}} 
\end{equation}
confirms the equality of the two projected spectra as a derived
result.



\end{document}